\documentclass{JHEP3} 

\usepackage{epsfig,multicol}
\usepackage{amsmath, amssymb}

\newcommand\fverb{\setbox\pippobox=\hbox\bgroup\verb}
\newcommand\fverbdo{\egroup\medskip\noindent%
			\fbox{\unhbox\pippobox}\ }
\newcommand\fverbit{\egroup\item[\fbox{\unhbox\pippobox}]}
\newbox\pippobox

\newcommand{\be}{\begin{equation}} 
\newcommand{\ee}{\end{equation}}

\newcommand{\ba}{\begin{eqnarray}}
\newcommand{\ea}{\end{eqnarray}}

\title{Bethe Ansatz solutions for highest states in ${\cal N}=4$ SYM and
AdS/CFT duality}

\author{Matteo Beccaria\thanks{Partially supported by INFN, IS-RM62}\\
        Dipartimento di Fisica, Universita' di Lecce,
        Via Arnesano, 73100 Lecce\\
        INFN, Sezione di Lecce\\
        E-mail: \email{matteo.beccaria@le.infn.it}}

\author{Luigi Del Debbio \\
        SUPA, School of Physics, University of Edinburgh,
        Edinburgh EH9 3JZ, UK\\
        E-mail: \email{luigi.del.debbio@ed.ac.uk}
}

\preprint{}

\abstract{ 
We consider the operators with highest anomalous dimension
$\Delta$ in the compact rank-one sectors $\mathfrak{su}(1|1)$ and
$\mathfrak{su}(2)$ of ${\cal N}=4$ super Yang-Mills.  We study the
flow of $\Delta$ from weak to strong 't Hooft coupling $\lambda$ by
solving (i) the all-loop gauge Bethe Ansatz, (ii) the quantum string
Bethe Ansatz.  The two calculations are carefully compared in the
strong coupling limit and exhibit different exponents $\nu$ in the
leading order expansion $\Delta\sim \lambda^{\nu}$.  We find $\nu =
1/2$ and $\nu = 1/4$ for the gauge or string solution. This strong
coupling discrepancy is not unexpected, and it provides an
explicit example where the gauge Bethe Ansatz solution cannot be
trusted at large $\lambda$.  Instead, the string solution perfectly
reproduces the Gubser-Klebanov-Polyakov law $\Delta =
2\sqrt{n}\,\lambda^{1/4}$. In particular, we provide an analytic
expression for the integer level $n$ as a function of the U(1) charge
in both sectors.  }


\begin{document} 

\section{Introduction}
\label{Sec:Intro}

The AdS/CFT
correspondence~\cite{Maldacena:1997re,Gubser:1998bc,Witten:1998qj,Kazakov:2004qf,Klebanov:2000me}
is a non-trivial map between two integrable theories, string theory on
$AdS_5\times S^5$ and the maximally supersymmetric ${\cal N}=4$ super
Yang-Mills SU(N) gauge theory (SYM).  In the planar limit
$N\to\infty$, the string coupling vanishes and the correspondence
relates a finite superconformal four dimensional theory and free
string theory on a non-trivial background.  Massive string states are
predicted to be dual to certain composite operators in the gauge
theory, with the string spectrum matching the gauge anomalous
dimensions. In terms of the planar 't Hooft coupling $\lambda =
g^2_{\rm YM}\,N$, the duality is of the weak-strong coupling type.
Hence, any test of the correspondence must exploit some kind of
non-perturbative knowledge on at least one of the two sides.

The large $\lambda$ limit is particularly interesting since the string
side can be controlled in the supergravity approximation.  A quite
general prediction is the scaling $E\sim 2\,\sqrt{n}\,\lambda^{1/4}$
for the energy of level $n$ massive string states as
$\lambda\to\infty$~\cite{Gubser:1998bc}. In the gauge theory, a check
of this prediction requires the knowledge of the anomalous dimensions
of suitable dual composite operators in the nonperturbative regime.
Such a formidable task is made possible by the integrability
properties of ${\cal N}=4$ SYM. The scaling operators are eigenstates
of the dilatation operator $\mathfrak{D}$ that can be identified with
the Hamiltonian of integrable (super) spin chains in various sectors
closed under perturbative renormalization. The integrable operator
$\mathfrak{D}$ can be treated by Bethe Ansatz
techniques~\cite{Minahan:2002ve,Beisert:2003jj,Beisert:2004ry}.  In
particular, all-loop conjectured gauge Bethe Ansatz (GBA) equations
are available in the compact $\mathfrak{su}(2)$, $\mathfrak{su}(1|1)$
and non-compact $\mathfrak{sl}(2)$
sectors~\cite{Staudacher:2004tk,Beisert:2005fw}.

Unfortunately, the GBA equations are only asymptotically exact. For
operators with classical dimension $L$, they predict the exact
anomalous dimension up to {\em wrapping terms} appearing at a certain
order increasing with $L$, for instance terms ${\cal O}(\lambda^L)$ in
the $\mathfrak{su}(2)$ sector~\cite{Beisert:2004hm}.  Due to wrapping
terms, the GBA equations are not reliable at strong coupling, although
in some cases they are believed to give the correct leading
term. Remarkably, in the $\mathfrak{su}(2)$ sector, a local version of
the GBA equations has been proposed in the form of a Hubbard-like
model~\cite{Rej:2005qt}; it has been conjectured to be free from
wrapping problems, but the reconciliation of its strong coupling
predictions with string theory is far from
clear~\cite{Rej:2005qt,Minahan:2006zi,Beccaria:2006aw}.

In general, the gauge and string calculations overlap in BMN-like
limits~\cite{Berenstein:2002jq} where $L$ is large. In this case, it
is well known that the perturbative comparison in powers of $\lambda$
is plagued by the different order of the limits $\lambda\to 0$,
$L\to\infty$ on the two sides. For instance, the exact three-loop
anomalous dimension of two- and three-magnon operators in the near-BMN
limit~\cite{Berenstein:2002jq} exhibits a three-loop discrepancy when
compared with the leading curvature correction computed in string
theory~\cite{Callan:2003xr,Callan:2004uv,Callan:2004ev,Callan:2004dt}.
Similar three-loop discrepancies also occur in the expansion around
spinning string solutions~\cite{Serban:2004jf,Beisert:2004hm}.

Along a different route, one can start from the classical string
theory (at large $\lambda$) and derive thermodynamical Bethe Ansatz
equations at $L\to\infty$. The discretization of these string Bethe
Ansatz equations (SBA) have been proposed to compute the leading $1/L$
effects, {\em i.e.} one-loop worldsheet quantum
corrections~\cite{Arutyunov:2004vx,Staudacher:2004tk,Beisert:2005fw}. The
validity of the SBA equations at finite $L$ or small $\lambda$ is not
guaranteed and indeed they are known to receive several kinds of
corrections~\cite{Beisert:2005cw,Schafer-Nameki:2005tn,Schafer-Nameki:2005is,Schafer-Nameki:2006gk}.
These corrections have been evaluated for various classes of
Frolov-Tseytlin spinning string
solutions~\cite{Frolov:2003qc,Frolov:2003xy,Arutyunov:2003uj,Arutyunov:2003za,Frolov:2003tu,Frolov:2004bh,Park:2005ji}.
They suggest the emergence of an interpolating set of Bethe Ansatz
equations working at all $\lambda$ and
$L$~\cite{Hernandez:2006tk,Freyhult:2006vr} and hopefully solving the
three-loop discrepancies.

In this paper, we take a complementary approach by comparing the GBA
and SBA equations at strong coupling.  Indeed, it is not totally clear
to what extent the GBA equations are able to predict the correct
results in the string regime $\lambda\to\infty$ as discussed for
instance in~\cite{Minahan:2006zi,Beccaria:2006aw,Arutyunov:2006av}.
We attempt to answer this question for the states with highest
anomalous dimension in the two compact rank-one $\mathfrak{su}(2)$ and
$\mathfrak{su}(1|1)$ subsectors where we are able to solve the GBA and
SBA equations at fixed $L$ and generic $\lambda$.  To appreciate the
special role of the highest states, we briefly summarize some relevant
facts about them.

\medskip
In the $\mathfrak{su}(2)$ sector, the highest state is the so-called
antiferromagnetic (AF) operator~\cite{Zarembo:2005ur}. It can be
defined in the multiplet of operators with fixed classical dimension
$L$. At the perturbative level, it mixes with the other states and its
explicit expression is not available in closed form at finite $L$.  In
the $L\to\infty$ limit, the GBA equations can be solved and give the
anomalous dimension 
\be
\label{eq:AF}
\lim_{L\to\infty} \frac{\Delta_{\mathfrak{su}(2)}}{L} =
 1+\frac{\sqrt{\lambda }}{\pi }\int_{0}^{\infty }
 \frac{ds}{s}\,\,\frac{J_0\left(\frac{\sqrt{\lambda }}{2\pi }s\right)
 J_1\left(\frac{\sqrt{\lambda }}{2\pi }s\right)}{\,{\rm e}\,^{s}+1}\ .
 \ee 
This expression is obtained by taking the $L\to\infty$ limit at fixed
 $\lambda$. Hence, it is legitimate to expand at weak coupling and one
 obtains
\be \lim_{L\to\infty} \frac{\Delta_{\mathfrak{su}(2)}}{L}
 =1+4\ln 2\,\frac{\lambda }{16\pi ^2}-9\,\zeta (3)\left(\frac{\lambda
 }{16\pi ^2}\right)^2 +75\,\zeta (5)\left(\frac{\lambda }{16\pi
 ^2}\right)^3+\dots .  
\ee 
On the other hand, the strong coupling expansion of Eq.~(\ref{eq:AF})
 is
\be
\label{eq:su2:strong}
\lim_{L\to\infty}\frac{\Delta_{\mathfrak{su}(2)}}{L} =
\frac{1}{\pi^2}\,\sqrt{\lambda} + \frac{3}{4} + \dots .  
\ee 
The same leading term is obtained in the Hubbard model formulation of
the GBA equations~\cite{Minahan:2006zi,Beccaria:2006aw}.  As discussed
in~\cite{Zarembo:2005ur} this expansion is formal due to the
uncontrolled effect of wrapping terms.  The usual attitude toward this
problem is rather optimistic and Eq.(\ref{eq:su2:strong}) with its
$\sim\lambda^{1/2}$ signature is expected to be correct apart from a
possible correction in the numerical prefactor $1/\pi^2$.  As a
support to the this scenario, it has been proposed to identify its
dual string state with a suitable string solution in the spirit of
similar correspondences found in the lowest part of the
spectrum~\cite{FrolovTseytlin}.  The {\em slow-string} solution
described in~\cite{Roiban:2006jt} exhibits the $\lambda^{1/2}$ scaling
of Eq.~(\ref{eq:su2:strong}) although with a different numerical
prefactor. This quantitative discrepancy has been attributed
in~\cite{Roiban:2006jt} to the subtle double limit $\lambda,
L\to\infty$.  However, the Hubbard model formulation~\cite{Rej:2005qt}
suggests that this scenario is not entirely satisfactory. There, limit
ambiguities and wrapping problems are absent and nevertheless the same
prediction $\Delta/L\sim \sqrt\lambda/\pi^2$ is recovered including
the prefactor~\cite{Minahan:2006zi,Beccaria:2006aw}.  A more
convincing proof should at least include the analysis of the solution
of the SBA equations, valid in the strong coupling limit.

A similar analysis can be attempted for the highest operator in the
other compact sector $\mathfrak{su}(1|1)$~\cite{Arutyunov:2006av}.
Here the precise form of the operator is known at finite $L$ and
simply reads $\mbox{Tr}(\psi^L)$ where $\psi$ is the highest weight
component of the Weyl spinor in the vector multiplet.  Unfortunately,
a closed formula like Eq.~(\ref{eq:AF}) is not known in this sector.
In~\cite{Arutyunov:2006av}, the GBA equations for the highest state
are studied at weak and strong coupling in the $L\to\infty$ limit.
The weak coupling expansion turns out to have a finite but rather
small convergence radius and is not immediately useful to reach the
strong coupling regime. On the other hand, the strong coupling
expansion is ambiguous and depends on assumptions about the large
$\lambda$ behavior of the Bethe parameters~\cite{Arutyunov:2004vx}.
Two asymptotic solutions for the Bethe momenta have been proposed
in~\cite{Arutyunov:2006av}. The leading term in the anomalous
dimension is the same in both cases and scales like $\lambda^{1/2}$.

To summarize, the present knowledge on the strong coupling behavior of
the maximal states in the compact sectors of ${\cal N}=4$ SYM is (we
write only the leading term at large $\lambda$ and define $L$ to be
the number of Bethe momenta in both sectors)
\be
\label{oldresults}
\begin{array}{ccll}
\displaystyle\lim_{L\to\infty}\frac{\Delta_{\mathfrak{su}(2)}^{\rm gauge}}{2L} &=& \displaystyle\frac{1}{\pi^2}\,\lambda^{1/2}, & (\mbox{exact})\\ \\
\displaystyle\frac{\Delta_{\mathfrak{su}(1|1)}^{\rm gauge}}{L} &=& \displaystyle c_L\,\lambda^{1/2}, & (\mbox{conjecture}\,\cite{Arutyunov:2006av}) 
\end{array}
\qquad
\begin{array}{ccl}
\displaystyle\lim_{L\to\infty}\frac{\Delta_{\mathfrak{su}(2)}^{\rm string}}{2L} &=& \mbox{?}, \\ \\
\displaystyle\lim_{L\to\infty}\frac{\Delta_{\mathfrak{su}(1|1)}^{\rm string}}{L} &=& \mbox{?}, 
\end{array}
\ee
$$
c_L \to \displaystyle \frac{3\sqrt{3}}{2\pi^2},\quad \mbox{as}\ L\to\infty, \nonumber
$$
\medskip
where the superscripts gauge/string label the results obtained with
the GBA/SBA equations.

In this paper, we pursue the analysis of these states with the aim of
filling the gaps in the above predictions.  We begin with the
$\mathfrak{su}(1|1)$ sector which is particularly favorable from the
technical point of view.  At weak coupling we present high-order
results for the anomalous dimension computed by the GBA equations.  We
show that a resummation is possible by a non-linear acceleration
method, the Weniger algorithm. It permits to evaluate the anomalous
dimension for rather large values of the coupling $\lambda$. This
leads to results supporting the $\lambda^{1/2}$ asymptotic behavior,
although with a different coefficient with respect to the proposals
in~\cite{Arutyunov:2006av}.

To investigate further the flow from weak to strong coupling we
present a numerical solution of the GBA equations. It is worthwile to
emphasize that the analysis is quite robust and can be extended to
very large values of $\lambda$ following in a clean way the evolution
of Bethe momenta. Our analysis reveals several subtleties involved in
the strong coupling expansion of the GBA equations.  The final result
is simple and we are able to compute very precisely the leading
strong-coupling term of the anomalous dimension. Not suprisingly, the
agreement with the weak coupling resummation is quite good. At this
point, the information about the highest state is similar to what is
known in the $\mathfrak{su}(2)$ sector.  We have accurately computed
the weak coupling expansion and the leading asymptotic $\lambda^{1/2}$
term, but we do not know to what extent the GBA are reliable in the
strong coupling region.  We remark that, in the $\mathfrak{su}(1|1)$
sector, we do not have a slow-string limit solution to be identified
with the highest state.

The next obvious step is to analyze the SBA equations in this sector
by the same methods. Once again the string Bethe Ansatz equations can
be integrated numerically. The result is very interesting. All
Bethe momenta flow to zero at large $\lambda$ with a simple leading
term $p_k\sim \alpha_k\,\lambda^{-1/4}$, and the asymptotic
coefficients $\alpha_k$ can be computed numerically. We also determine
analytically the prefactor in the leading term $\sim \lambda^{1/4}$ in
the anomalous dimension.

The same analysis can be applied to the $\mathfrak{su}(2)$ sector. The
weak coupling resummation does not apply here because the precise form
of the highest AF state depends on $L$. However, the numerical and
analytical study of the strong-coupling behavior of GBA and SBA
equations can be performed without difficulty. We find a pattern
similar to the $\mathfrak{su}(1|1)$ one.  Our results are summarized
in the following table which has to be compared with
Eqs.~(\ref{oldresults}).  In the right hand sides, we only report the
leading term at large $\lambda$ and finite $L$ 
\be
\label{result}
\begin{array}{ccl}
\displaystyle\frac{\Delta_{\mathfrak{su}(2)}^{\rm gauge}}{2L}  &=& \displaystyle \frac{1}{2\pi L\sin{\displaystyle\frac{\pi}{2L}}}\,\lambda^{1/2},\\ \\
\displaystyle \frac{\Delta_{\mathfrak{su}(1|1)}^{\rm gauge}}{L}  &=& \displaystyle c_L\,\lambda^{1/2}, 
\end{array}
\qquad
\begin{array}{ccl}
\displaystyle \frac{\Delta_{\mathfrak{su}(2)}^{\rm string}}{2L} &=&  \displaystyle\frac{1}{2}\,\lambda^{1/4}, \\ \\
\displaystyle \frac{\Delta_{\mathfrak{su}(1|1)}^{\rm string}}{L} &=& \displaystyle \frac{1}{\sqrt{2}}\left(1-\frac{1}{L^2}\right)\, \lambda^{1/4},
\end{array}
\ee
$$
c_L \to   0.1405(1),\quad \mbox{as}\ L\to\infty.
$$
\medskip

The ratio $\Delta_{\mathfrak{su}(2)}^{\rm string}/(2L)$ is
independent on $L$. Also, in the $L\to\infty$ limit, we recover the
exact prefactor $1/\pi^2$ for $\Delta_{\mathfrak{su}(2)}^{\rm
gauge}/(2L)$.

\medskip

The physical contents of Eqs.~(\ref{result}) will be discussed in
Sec.~\ref{sec:Discussion}, after heving illustrated the technical
details of the derivation. These will be organized as follows.  In
Sec.~\ref{sec:GBA11}, we present the gauge Bethe Ansatz equations for
the $\mathfrak{su}(1|1)$ sector.  In Sec.~\ref{sec:GBA11weak} and
\ref{sec:GBA11strong} we analyze them for the highest state obtaining
in particular the strong coupling expansion of the anomalous
dimension.  Sec.~\ref{sec:SBA11} presents the string Bethe Ansatz
equations, again for the $\mathfrak{su}(1|1)$ sector and
Sec.~\ref{sec:SBA11weak} and \ref{sec:SBA11strong} repeat the previous
analysis. In this case, the strong coupling expansion of the anomalous
dimension is determined exactly.  In Sec.~\ref{sec:GBA2} and
\ref{sec:SBA2} we present a similar analysis for the other compact
$\mathfrak{su}(2)$ sector.

\section{The gauge Bethe Ansatz equations for the highest state in the $\mathfrak{su}(1|1)$ sector}
\label{sec:GBA11}

The dilatation operator in the $\mathfrak{su}(1|1)$ sector can be
associated with a super spin
chain~\cite{Beisert:2003ys,Callan:2004dt}. The all-loop gauge Bethe
Ansatz equations have been proposed
in~\cite{Staudacher:2004tk,Beisert:2005fw} and read
\be
e^{i\,L\,p_k} = \prod_{j\neq k}\frac{\displaystyle 1-\frac{g^2}{2\,x^+(p_k)\,x^-(p_j)}}
{\displaystyle 1-\frac{g^2}{2\,x^-(p_k)\,x^+(p_j)}},\qquad k = 1, \dots, L,
\ee
where $L\in 2 \mathbb{N}+1$ and 
\be
x^\pm(p) = \frac{e^{\pm i\,\frac{p}{2}}}{4\sin\frac{p}{2}}\left(1+\sqrt{1+8\,g^2\,\sin^2\frac{p}{2}}\right).
\ee
The coupling $g$ is related to the 't Hooft coupling by $\lambda =
8\,\pi^2\, g^2$.  The anomalous dimension of the state associated with
the solution $\{p_k(g^2)\}$ is
\be
\Delta = \frac{3}{2}\,L + \sum_{k=1}^L\left(\sqrt{1+8\,g^2\,\sin^2\frac{p_k}{2}}-1\right).
\ee
The Bethe momenta of the highest state at $g=0$ are
\be
\label{eq:zero}
p_k = \frac{2\pi}{L}\,n_k,\qquad n_k = -\frac{L-1}{2}, \dots, \frac{L-1}{2}.
\ee
The GBA equations in logarithmic forms are
\be
p_k = \frac{2\pi}{L}\,n_k -\frac{i}{L}\sum_{j\neq k}
\log\frac{\displaystyle 1-\frac{g^2}{2\,x^+_k\,x^-_j}}
{\displaystyle 1-\frac{g^2}{2\,x^-_k\,x^+_j}},
\ee
where $n_k$ are given in (\ref{eq:zero}). They are suitable for weak
coupling expansions since the second term in the r.h.s. is ${\cal
O}(g^2)$.

\subsection{Weak coupling expansion and Weniger resummation}
\label{sec:GBA11weak}

The weak-coupling expansion of the anomalous dimension is easily
obtained. We simply start with the zero-th order value of Bethe
momenta for a certain fixed $L$, Eq.~(\ref{eq:zero}).  Then, we
replace them in the GBA equation and expand the r.h.s. at first order
in $g^2$. Repeating this procedure and expanding the expression for
$\Delta$ order by order, we obtain the perturbative expansion of
$\Delta$.

In the ratio $\Delta/L$, the terms up to ${\cal O}(g^{2L-2})$ do not
change if $L$ is increased. For instance,
\ba
\left(\frac{\Delta}{L}\right)_{L=3} &=& \frac{3}{2}+2\,g^2-4\,g^4+14\,g^6-\frac{235}{4}\,g^8+\frac{2209}{8}\,g^{10} + \cdots, \\
\left(\frac{\Delta}{L}\right)_{L=5} &=& \frac{3}{2}+2\,g^2-4\,g^4+\frac{29}{2}\,g^6-\frac{259}{4}\,g^8+\frac{2611}{8}\,g^{10} + \cdots, \\
\left(\frac{\Delta}{L}\right)_{L=7} &=& \frac{3}{2}+2\,g^2-4\,g^4+\frac{29}{2}\,g^6-\frac{259}{4}\,g^8+\frac{1307}{4}\,g^{10} + \cdots, 
\ea
This remark allows one to compute the $L\to\infty$ limit of the
expansion at a fixed order in $g$ by simply taking a sufficiently
large $L$. The procedure can be performed at the semi-numerical
level. In other words, we work with finite high precision using
numerical values of the momenta in a symbolic algebra calculation.  If
the precision is suitably high, the identification of the coefficients
of the energy expansion can be unambiguously identified with rational
numbers. We have performed the calculation up to the term
$g^{68}$. This requires $L\ge 35$. The result is
\ba
\lim_{L\to\infty}\frac{\Delta}{L}  &=&
\textstyle
\frac 3 2 + 2\,g^2 - 4\,g^4 + \frac{29}{2}\,g^{6} - \frac{259}{4}\,g^{8} + \frac{1307}{4}\,g^{10} - 
  1790\,g^{12} + 10396\,g^{14} - \frac{504397}{8}\,g^{16} \nonumber \\
&&\textstyle + \frac{6324557}{16}\,g^{18} - 
  \frac{40702709}{16}\,g^{20} + \frac{8561442701}{512}\,g^{22} - \frac{114529021311}{1024}\,g^{24} \nonumber \\
&&\textstyle + 
  \frac{777307887947}{1024}\,g^{26} - \frac{2670717561365}{512}\,g^{28} + \frac{37098574647961}{1024}\,g^{30} - 
  \frac{4161069724993527}{16384}\,g^{32}\nonumber \\
&&\textstyle  + \frac{29408079892945107}{16384}\,g^{34} - 
  \frac{104670245742870895}{8192}\,g^{36} + \frac{5999052730939686071}{65536}\,g^{38} \nonumber \\
&&\textstyle  - 
  \frac{86452214868942845981}{131072}\,g^{40} + \frac{626162974135003430373}{131072}\,g^{42} - 
  \frac{4556537471418865642837}{131072}\,g^{44}\nonumber \\
&&\textstyle  + \frac{66598702591887298874029}{262144}\,g^{46} - 
  \frac{244303906058015917431755}{131072}\,g^{48}\nonumber \\
&&\textstyle  + \frac{7195137546781961772111605}{524288}\,g^{50} - 
  \frac{106303776929607820974312023}{1048576}\,g^{52}\nonumber \\
&&\textstyle  + \frac{49230031886653815687152661}{65536}\,g^{54} - 
  \frac{45726319914455572899079305}{8192}\,g^{56}\nonumber \\
&&\textstyle  + \frac{348843198908576206971428650203}{8388608}\,g^{58} - 
  \frac{2605279742772089252587976183821}{8388608}\,g^{60}\nonumber \\
&&\textstyle  + \frac{39003030225010830621366145740085}{16777216}\,g^{62} - 
  \frac{2340608578131628813286501122058923}{134217728}\,g^{64}\nonumber \\
&&\textstyle  + \frac{35185861176795745832756768610959237}{268435456}\,g^{66}\nonumber\\
&&\textstyle  - 
  \frac{264968465576189708105542064159612145}{268435456}\,g^{68} + {\cal O}(g^{70}). 
\ea
We have identified the rational numbers by working with 200 digits
arithmetics~\footnote{For instance, with such a precision, the
coefficient of $g^{68}$ appears in the calculation as a floating point
number $r$ such that
\be
268435456\,r = 264968465576189708105542064159612145.\underbrace{000\cdots 0}_{163\,\,\rm null\ digits},
\ee
allowing for a safe identification of the numerator of the rational
coefficient. Standard packages like {\tt Mathematica}~\cite{Math}
allows easily this kind of high precision numerical calculations.  }.
This power series is convergent for $g^2 \lesssim \frac{1}{8}$, a
rather small convergence radius.  It is definitely useless to evaluate
strong coupling behavior at least in this form. We need an analytical
continuation beyond the convergence radius. Since the series is
alternating, we have attempted such continuation by means of the
non-linear Weniger algorithm~\cite{Weniger} that we describe in
Appendix~\ref{app:weniger}.

In our case, the Weniger algorithm is found to work very well. As an
example, we consider $g^2=1$ which is far beyond the convergence
radius. The Weniger approximants are shown in the first curve of
Fig.~(\ref{fig:wenigerconvergence}) where a clear and definite
convergence is achieved.  Going to higher values of the coupling, we
find that the resummation algorithm gives stable results up to
$g^2\simeq 20$, which is a fairly high value.  The other two curves of
Fig.~(\ref{fig:wenigerconvergence}) shows the behavior of Weniger
approximants for $g^2=10, 30$.

The plot of $\Delta/L-3/2$ at $L\to\infty$ in the stability region is
shown in Fig.~(\ref{fig:weniger}) where we also draw the results from
the numerical analysis of the GBA equations that are discussed in the
next Section.  From the Weniger algorithm, we recover clearly the
asymptotic $\sqrt{\lambda}$ behavior. A numerical fit gives the
estimate
\be
\lim_{L\to\infty}\frac{\Delta}{L}\sim 0.1404\,\sqrt{\lambda}.
\ee
The numerical prefactor is different than the one predicted
in~\cite{Arutyunov:2006av}. The question is whether the resummation is
failing or the strong coupling expansion is revealing some
surprise. In the next Section, we shall answer this question in favor
of the second hypothesis.

\subsection{Numerical solution and strong coupling behavior}
\label{sec:GBA11strong}

\subsubsection{Preliminary remarks}

The form of the GBA equations at strong coupling depends crucially on
certain {\em a priori} assumptions about the asymptotic form of Bethe
momenta. For the highest state, it is necessary to consider separately
three different cases, motivated by the following numerical
analysis. If we denote by $p(g)$ a particular running Bethe momentum,
we consider the three special large-$g$ behaviors
\ba
\mbox{I} &:& p(g)\to \overline p>0, \nonumber \\
\mbox{II} &:& p(g)\sim \alpha\,g^{-1/2}, \\
\mbox{III} &:& p(g)\sim  \alpha\,g^{-1}. \nonumber 
\ea
The ratio $g/x^\pm(p)$ has the following limit
\be
\label{eq:limits}
\frac{g}{x^\pm(p)} \to \left\{
\begin{array}{ll}
\mbox{I:} & \sqrt{2}\,e^{\mp i \frac{\overline p}{2}}\,\varepsilon(\sin\frac{\overline p}{2}), \\ \\
\mbox{II:} & \sqrt{2}\,\varepsilon(\alpha), \\ \\
\mbox{III:} & \displaystyle \frac{2 \alpha}{1+\sqrt{1+2\alpha^2}},
\end{array}
\right.
\ee
where $\varepsilon(x) = x/|x|$ is the sign function. For better
uniformity, it is convenient to write the case II as
\be
\sqrt{2}\,\varepsilon(\alpha) \equiv \sqrt{2}\,e^{\mp i \frac{\overline p}{2}}\,\varepsilon(\sin\frac{\overline p}{2}), 
\ee
where $\overline p=0$ in this case, and $\varepsilon(0)=\pm 1$
according to the sign of $\alpha$. With these conventions, cases I and
II are expressed by the same formula.

\subsubsection{The Arutyunov-Tseytlin Ansatz}

An Ansatz for the strong coupling behavior of the Bethe momenta of the
highest state is described in~\cite{Arutyunov:2006av}.  As we shall
discuss, it is closely related to the actual solution. The
Arutyunov-Tseytlin Ansatz assumes that one $p$ remains zero and the
other tend to non-zero limits symmetrically distributed around
zero. This Ansatz is quite reasonable since the symmetric pattern is
valid at $g=0$ and remains true at all orders in the weak coupling
expansion.

Under this assumption, we can look at the positive $p$ only. Using the
expressions~(\ref{eq:limits}), the GBA equation for any of them
reduces at strong coupling to
\be
e^{iLp_k} = e^{-i p_k}\, \mathop{\prod_{j=1}^{\frac{L-1}{2}}}_{j\neq k}
\frac{1-e^{-\frac{i}{2}(p_k-p_j)}}{1-e^{\frac{i}{2}(p_k-p_j)}}\,
\frac{1+e^{-\frac{i}{2}(p_k+p_j)}}{1+e^{\frac{i}{2}(p_k+p_j)}}
= -\prod_{j=1}^{\frac{L-1}{2}} (-e^{-i p_k})
\ee
This gives
\be
L p_k = \frac{L-3}{2}\pi-\frac{L-1}{2} p_k + 2\pi m_k,\qquad m_k\in\mathbb{Z},
\ee
or
\be
\label{AT:Ansatz}
p_k = \frac{4\pi\,m_k}{3L-1} + \pi\frac{L-3}{3L-1}.
\ee

\subsubsection{Explicit results at various $L$}

As we discussed, the result~(\ref{AT:Ansatz}) implies an asymptotic
anomalous dimension which does not agree with the resummation
results. To understand what is happening, we have solved numerically
the gauge Bethe Ansatz equations according to the following recipe
\begin{enumerate}
\item we start with the solution at $g=0$,
\item we progressively increase $g$ and solve step by step the Bethe
equations by Newton's algorithm~\cite{Newton},
\item at each step, we use the solution at the previous $g$ as a
starting guess for Newton's algorithm.
\end{enumerate} 
The procedure turns out to be very stable and can be extended up
to very large $\lambda$ values.  In particular it is possible to
increase $g$ in logarithmic scale. The stability of the algorithm is
checked by varying the numerical precision used in the intermediate
computations. We never encountered any singularity.  By means of this
numerical method, we have investigated the GBA equations at several
$L$ in order to discover why and when the Ansatz~(\ref{AT:Ansatz})
fails.
\medskip

Data for $L=3$ and $L=5$ are reported in Fig.~(\ref{fig:L=3:p}); they
confirm the Ansatz (\ref{AT:Ansatz}). We show the positive Bethe
momentum and the prediction with $k=1$ for $L=3$, and $k=0,2$ for
$L=5$.  Notice that convergence is achieved at quite large $\lambda$.

At $L=7$, the Bethe momenta are shown in Fig.~(\ref{fig:L=7:p}). Here,
something new happens. One of the Bethe momenta tends to zero.
Nevertheless, the Ansatz (\ref{AT:Ansatz}) is still working, with
$k=0, 2$. The reason is that the vanishing momentum tends to zero like
$\lambda^{-1/4}$ (case II) and the limiting form of the GBA equations
is the same as it would be in case I. The asymptotic form of the
vanishing momentum is illustrated in Fig.~(\ref{fig:L=7:pvanishing}).

At $L=9$, the Bethe momenta are shown in Fig.~(\ref{fig:L=9:p}). Here
again one of the momenta tends to zero.  Now, the Ansatz
(\ref{AT:Ansatz}) fails to predict the correct asymptotic values of
the non-vanishing momenta.  Indeed, the vanishing momentum tends to
zero like $\lambda^{-1/2}$ (case III) as shown in
Fig.~(\ref{fig:L=9:pvanishing}) and the limiting form of the GBA
equations is changed. The dashed lines predicting the actual
asymptotic non-zero $p$ are obtained as follows. Using again
Eqs.~(\ref{eq:limits}), the GBA equation for any positive $p$ reads in
the strong coupling limit ($L=9$)
\be
\label{eq:comment1}
e^{iLp} = e^{-3\,i\,p}\frac{1-\rho^2\, e^{-ip}}{1-\rho^2\, e^{ip}},\qquad \rho = \frac{\sqrt{2}\,\alpha}{1+\sqrt{1+2\alpha^2}},
\ee
where $\alpha$ appears in the asymptotic form of the vanishing
momentum which is $\alpha/\sqrt\lambda$.  For each $\rho$ we can
determine the three positive $p$ nearest to the numerical asymptotic
values. Then, we fix the parameter $\rho$ by using the strong coupling
limit of the GBA equation for the vanishing momentum
\be
\label{eq:comment2}
1 = \prod_{j=1, 2, 3}\frac{1-\rho\, e^{i\frac{p_j}{2}}}{1-\rho\, e^{-i\frac{p_j}{2}}}
\frac{1+\rho\, e^{-i\frac{p_j}{2}}}{1+\rho\, e^{i\frac{p_j}{2}}},
\ee
The numerical solution is easily found. With 25 digits, it reads
\ba
\rho &=& 0.7261948032180057677773276, \\
p_1  &=& 0.6047720145641787805731663, \nonumber \\
p_2  &=& 1.648738996485669279031225, \nonumber \\
p_3  &=& 2.646257150202204974776960.\nonumber 
\ea
The agreement, shown in Fig.~(\ref{fig:L=9:p}) is excellent.  In the
following of this paper, we shall often have to solve equations like
Eqs.~(\ref{eq:comment1}-\ref{eq:comment2}).  Whenever we claim that
{\em a numerical solution is easily found}, we mean that standard
packages, like {\tt Mathematica}~\cite{Math}, can determine the
solution with high precision in a straightforward way. For simplicity,
we shall give just a relatively small number of digits for such
results, but in all cases, we have checked their stability by
increasing the precision and checking that the result is unchanged.

At $L=11$, the Bethe momenta are shown in
Fig.~(\ref{fig:L=11:p}). Here again one of the momenta tends to zero
like $\lambda^{-1/2}$ (case III) and the limiting form of the GBA
equations is changed.  As before, we can compute the dashed lines
showing the asymptotic non-zero $p$. The GBA equations for the
positive $p$ read ($L=11$)
\be
e^{iLp} = -e^{-4\,i\,p}\frac{1-\rho^2\, e^{-ip}}{1-\rho^2\, e^{ip}},
\ee
Again, for each $\rho$ we can determine the four positive $p$ nearest
to the numerical asymptotic values. Then, $\rho$ is fixed by the GBA
equation for the vanishing momentum which reads
\be
1 = \prod_{j=1, 2, 3, 4}\frac{1-\rho\, e^{i\frac{p_j}{2}}}{1-\rho\, e^{-i\frac{p_j}{2}}}
\frac{1+\rho\, e^{-i\frac{p_j}{2}}}{1+\rho\, e^{i\frac{p_j}{2}}},
\ee
The solution is now
\ba
\rho &=& 0.616048, \\
p_1  &=& 0.227432, \nonumber \\
p_2  &=& 1.09891,  \nonumber \\
p_3  &=& 1.92557, \nonumber \\
p_4  &=& 2.73741. \nonumber 
\ea
The agreement is shown in Fig.~(\ref{fig:L=11:p}).

At $L=13$, the Bethe momenta are shown in
Fig.~(\ref{fig:L=13:p}). Here two momenta tend to zero, one like
$\lambda^{-1/2}$ and the other like $\lambda^{-1/4}$. We repeat the
exercise of computing the asymptotic $p$.  The GBA equation for the
positive $p$ reads ($L=13$)
\be
e^{iLp} = e^{-5\,i\,p}\frac{1-\rho^2\, e^{-ip}}{1-\rho^2\, e^{ip}},
\ee
For each $\rho$ we can determine the four positive $p$ nearest to the
numerical asymptotic values. Then, we fix the parameter $\rho$ as
before by the GBA equation for the vanishing momentum which reads
\be
1 = \prod_{j=1, 2, 3, 4}\frac{1-\rho\, e^{i\frac{p_j}{2}}}{1-\rho\, e^{-i\frac{p_j}{2}}}
\frac{1+\rho\, e^{-i\frac{p_j}{2}}}{1+\rho\, e^{i\frac{p_j}{2}}},
\ee
The solution is 
\ba
\rho &=& 0.53244, \nonumber \\
p_1  &=& 0.72414, \nonumber \\
p_2  &=& 1.42787, \\
p_3  &=& 2.11751, \nonumber \\
p_4  &=& 2.80082. \nonumber
\ea
The agreement is shown in Fig.~(\ref{fig:L=13:p}).

If $L$ is further increased, the pattern of vanishing and non
vanishing momenta turns out to be quite regular.  In the following
table we show for each $L$ the number $N_{1/4}$ of positive momenta
vanishing like $\lambda^{-1/4}$ and the number $N_{1/2}$ of those
vanishing like $\lambda^{-1/2}$.
\TABLE{
\begin{tabular}{|| c|ccc|ccc|ccc|ccc|c || }
\hline
$L$             & 7 & 9 & 11 & 13 & 15 & 17 & 19 & 21 & 23 & 25 & 27 & 29 & $\cdots$\\
$N_\frac{1}{2}$ & 0 & 1 &  1 &  1 &  2 &  2 &  2 &  3 &  3 &  3 &  4 & 4  & $\cdots$\\
$N_\frac{1}{4}$ & 1 & 0 &  0 &  1 &  0 &  0 &  1 &  0 &  0 &  1 &  0 & 0  & $\cdots$\\
\hline
\end{tabular}
\caption{Periodicity of the number of vanishing Bethe momenta.}
}
The general formulas expressing the $\mathbb{Z}_3$ regularity of the Table are
\ba
N_\frac{1}{2} &=& \lfloor \frac{L-1}{6} \rfloor-1 + \left\{\begin{array}{ll}
0, & \displaystyle \frac{L-1}{2}\, \mbox{mod}\,3 = 0, \\ \\
1, & \mbox{otherwise}
\end{array}
\right. \\ 
\nonumber \\
N_\frac{1}{4} &=& \left\{\begin{array}{ll}
1, & \displaystyle \frac{L-1}{2}\, \mbox{mod}\,3 = 0, \\ \\
0, & \mbox{otherwise}
\end{array}
\right.
\ea
For instance, if $L=43$ we expect 6 momenta vanishing like
$\lambda^{-1/2}$ and one like $\lambda^{-1/4}$. The full set of
momenta is shown in Fig.~(\ref{fig:L=43:p}). It can be checked that
the vanishing momenta have precisely these asymptotic behaviors.

\subsubsection{Asymptotic form of $\Delta$}

The asymptotic form of the anomalous dimension at fixed $L$ and large
$\lambda$ is
\be
\frac{\Delta}{L} \sim c_L\,\sqrt{\lambda}\qquad {\rm with}\ c_L = \frac{1}{L\,\pi}\sum_{k=1}^L \left|\sin\frac{p_k}{2}\right|,
\ee
where the $p_k$ are the asymptotic non-zero values of the Bethe
momenta.  Due to the above periodicity, we can estimate $c_\infty$ by
considering separately our data for $c_L$ for the three values of
$((L-1)/2)\,{\rm mod}\,3$. The result from a fit of the data at $L>5$
by using a cubic polynomial in $1/L$ are shown in
Fig.~(\ref{fig:cfit}).  The three subsequences have clearly the same
limit. We find
\be
c_\infty = 0.1405(1),
\ee
where the error is a conservative estimate of the finite $L$ fit.

\medskip
A remark about the order of the two limits $L, \lambda\to \infty$ is
in order. We applied the resummation algorithm to estimate the leading
term at large $\lambda$ of $\lim_{L\to\infty}\Delta/L$. Here, solving
the GBA equations, we are fixing $L$ and taking the large $\lambda$
leading term $\sim c_L\sqrt{\lambda}$. Then, we evaluate $c_\infty =
\lim_{L\to\infty} c_L$.  Therefore, the double limit $L, \lambda\to
\infty$ is taken in two different orders. Nevertheless, the agreement
of the leading term in the two calculations is not surprising. This is
precisely what happens for the AF state in the $\mathfrak{su}(2)$
sector.  There, one can start from Eq.~(\ref{eq:AF}) and take after
the $\lambda\to\infty$ limit. Alternatively, one can take the large
$\lambda$ limit at fixed $L$, {\em e.g.} in the Hubbard model
formulation. The result for the first three terms in the expansion is
the same as discussed in~\cite{Beccaria:2006aw}.

\section{The string Bethe Ansatz equations for the highest state in the $\mathfrak{su}(1|1)$ sector}
\label{sec:SBA11}

The analysis of the GBA equations is certainly interesting, but the
ultimate goal is the comparison with string theory.  As discussed in
the Introduction, it is not clear to what extent the GBA predicts
correct results at strong coupling.  This question can be investigated
by studying the string Bethe Ansatz
equations~\cite{Staudacher:2004tk,Beisert:2005fw} expected to predict
the correct strong coupling behavior of string states, at least at
large $L$.

In order to write the SBA equations in a compact way, we define
$x^\pm_k = x^\pm(p_k)$ and $u_k=u(p_k)$ where 
\be u(p) = \frac 1 2
\cot\frac{p}{2}\sqrt{1+8\,g^2\,\sin^2\frac{p}{2}}.  \ee The string
Bethe Ansatz equations are then \be e^{i\,L\,p_k} = \prod_{j\neq k}
\frac{\displaystyle 1-\frac{g^2}{2\,x^+_k\,x^-_j}} {\displaystyle
1-\frac{g^2}{2\,x^-_k\,x^+_j}}\,e^{i\vartheta(p_k, p_j)}, 
\ee 
where the scattering phase $\vartheta$ is
\be
\label{eq:scattering}
e^{i\vartheta(p_k, p_j)} = \left(
\frac{\displaystyle 1-\frac{g^2}{2\,x^+_k\,x^-_j}}
{\displaystyle 1-\frac{g^2}{2\,x^-_k\,x^+_j}}
\right)^{-2}\,\,
\left(
\frac{\displaystyle 1-\frac{g^2}{2\,x^+_k\,x^-_j}}
{\displaystyle 1-\frac{g^2}{2\,x^-_k\,x^-_j}}
\,
\frac{\displaystyle 1-\frac{g^2}{2\,x^-_k\,x^+_j}}
{\displaystyle 1-\frac{g^2}{2\,x^+_k\,x^+_j}}
\right)^{2\,i\,(u_k-u_j)}.
\ee
In logarithmic form we have 
\be
p_k = \frac{2\pi}{L}\,n_k -\frac{i}{L}\sum_{j\neq k}\left(
-\log\frac{\displaystyle 1-\frac{g^2}{2\,x^+_k\,x^-_j}}
{\displaystyle 1-\frac{g^2}{2\,x^-_k\,x^+_j}}
+2\,i\,(u_k-u_j)\log \frac{\displaystyle 1-\frac{g^2}{2\,x^+_k\,x^-_j}}
{\displaystyle 1-\frac{g^2}{2\,x^-_k\,x^-_j}}
\,
\frac{\displaystyle 1-\frac{g^2}{2\,x^-_k\,x^+_j}}
{\displaystyle 1-\frac{g^2}{2\,x^+_k\,x^+_j}}
\right).
\ee
These equations are considerably more involved than the GBA
ones. Nevertheless, we have been able to repeat step by step the
previous analysis as we now discuss.

\subsection{Weak coupling expansion and Weniger resummation}
\label{sec:SBA11weak}

We repeat the semi-numerical algorithm that we followed for the weak
coupling expansion of the GBA equations.  We take again $L=35$ and
obtain the result
\ba
\lefteqn{\lim_{L\to\infty}\frac{\Delta^{\rm string}}{L} = 
\textstyle
\frac{3}{2} + 2\,g^{2} - 4\,g^{4} + \frac{25}{2}g^{6} - \frac{601}{12}g^{8} + 
  \frac{2849}{12}g^{10} - \frac{25141}{20}g^{12} + \frac{429809}{60}g^{14} - 
  \frac{9022721}{210}g^{16}} \nonumber \\
&&\textstyle + \frac{149821573}{560}g^{18} - \frac{8640293477}{5040}g^{20} + 
  \frac{1812303079883}{161280}g^{22} - \frac{88730558092937}{1182720}g^{24}\nonumber \\
&&\textstyle + 
  \frac{1804497110708207}{3548160}g^{26} - \frac{26846650998855167}{7687680}g^{28} + 
  \frac{279571052498891591}{11531520}g^{30}\nonumber \\
&&\textstyle - \frac{125402745492098095339}{738017280}g^{32} + 
  \frac{9738744677918359729}{8110080}g^{34} - \frac{6703245537745284313789}{784143360}g^{36}\nonumber \\
&&\textstyle  + 
  \frac{1024647660942740023729097}{16728391680}g^{38} - 
  \frac{2525524043347946614344579101}{5721109954560}g^{40} \nonumber \\
&&\textstyle + 
  \frac{18296543439438265154466562553}{5721109954560}g^{42} - 
  \frac{1331791971895366969823043659509}{57211099545600}g^{44}\nonumber \\
&&\textstyle  + 
  \frac{9735683703553399316563746438049}{57211099545600}g^{46} - 
  \frac{345964047352595964568301174322841}{277022166220800}g^{48} \nonumber \\
&&\textstyle + 
  \frac{48414229366853228412090322467734657}{5263421158195200}g^{50} - 
  \frac{2689943057205158649159985079500991}{39574595174400}g^{52}\nonumber \\
&&\textstyle  + 
  \frac{5303581496963536951888629403691895353}{10526842316390400}g^{54}\nonumber \\
&&\textstyle  - 
  \frac{19710992154318324101213041165671461761}{5263421158195200}g^{56}\nonumber \\
&&\textstyle  + 
  \frac{870504639734489849012829744121725553}{31190643900416}g^{58}\nonumber \\
&&\textstyle  - 
  \frac{10693434560671288576899374290035763214898727}{51286775765454028800}g^{60}\nonumber \\
&&\textstyle  + 
  \frac{76039980287752685716954083401449551327779}{48705390090649600}g^{62}\nonumber \\
&&\textstyle - 
  \frac{10643182102443547953784209313106048717253607517}{908508599273757081600}g^{64} \nonumber \\
&&\textstyle + 
  \frac{2240647681205846470476844290367422504232816365029}{25438240779665198284800}g^{66}\nonumber \\ 
&&\textstyle  - 
  \frac{5626138712126728829417020286414076511813854415733}{8479413593221732761600}g^{68} + {\cal O}(g^{70}).
\ea
The Weniger resummation algorithm is convergent for $g^2\lesssim
10$. We show the resummed expression for $\Delta$ in the left panel of
Fig.~(\ref{fig:compare}) for the gauge and string cases. The right
panel shows the derivative
\be
\frac{d}{d\log\lambda} \log\left(\lim_{L\to\infty}\frac{\Delta^{\rm string}}{L}\right),
\ee
which estimates at large $\lambda$ the exponent of the leading term.
In the gauge case, it approaches the value $1/2$ at large $\lambda$,
as we discussed.  In the string case, the asymptotic value appears to
be definitely smaller and the figure is qualitatively compatible with
an asymptotic behavior $\sim \lambda^{1/4}$.

To support this conclusion, we have fitted the whole data with the
functional form
\be
\lim_{L\to\infty}\frac{\Delta^{\rm string}(\lambda)}{L} = c_0\,\lambda^\nu + c_1 + c_2\,\lambda^{-\nu} + c_3\,\lambda^{-2\nu}.
\ee
The standard $\chi^2$ is a quantitative measure of the deviation from
the supposed dependence on $\lambda$. The best values for the exponent
$\nu$ are
\be
\mbox{gauge BA}\,:\, \nu_{\rm fit} = 0.496,\quad
\mbox{string BA}\,:\, \nu_{\rm fit} = 0.24,
\ee
which are very close to $1/2$ and $1/4$. If we now fix the exponent
$\nu=1/2$ or $1/4$, we find the following $\chi^2$ for the two curves:
\be
\begin{array}{ccc}
\hline
\hline 
\mbox{fixed exponent} & \nu = 1/2 & \nu = 1/4 \\ 
\hline
\hline \\
\chi^2(\mbox{gauge BA})  & 2.3\cdot 10^{-5} & 0.02 \\ \\
\chi^2(\mbox{string BA}) & 0.03 & 1.4\cdot 10^{-4} \\ \\
\hline
\hline
\end{array}
\ee
\medskip

As a conclusion, the resummed anomalous dimension favors the choice
$\nu=1/4$ in the string case. The leading term with its numerical
prefactor is
\be
\lim_{L\to\infty}\frac{\Delta^{\rm string}}{L} = 0.70(1)\,\lambda^{1/4}.
\ee
The prefactor is difficult to estimate and a better determination
would require a stable resummation at larger $\lambda$.

While this result is quite pleasing, it must be criticized because of
the moderate resummation range. Based on the weak coupling arguments
[presented so far, it can not be ecluded that the string curve in the
right panel of Fig.~(\ref{fig:compare}) could rise at larger $\lambda$
and flow back to the gauge value $1/2$.  To pursue the analysis, as in
the gauge case, we turn to a numerical iterative solution of the SBA
equations. Indeed, it should be clear that several solutions are
possible at strong coupling and the problem is again that of choosing
the right one.

\subsection{Exact solution at strong coupling}
\label{sec:SBA11strong}

We now determine the numerical solution of the SBA equations. We
follow the same procedure we described for the gauge BA equations.
The result is fully consistent with the weak coupling resummation: All
Bethe momenta $p_k$ vanish at large $\lambda$ with an asymptotic
behavior $p_k\sim \alpha_k\,\lambda^{-1/4}$ for all $p_k$!  We
illustrate this noticeable result by showing in Fig.~(\ref{fig:SBA})
the evolution of Bethe momenta scaled by $\lambda^{1/4}$ in the four
cases $L=3, 5, 15, 29$.

The coefficients $\alpha_k$ are symmetrically distributed around zero.
Qualitatively, $L-2$ coefficients $\alpha_k$ are {\em almost} evenly
spaced around zero. Two special momenta have instead coefficients
$\alpha_k$ well separated from the central band. Looking in more
details at the explicit solution, one finds that the $\alpha_k$ in the
central band have a non-trivial asymptotic density. The analysis of
the large $L$ form of this density if deferred to future work.

It is not difficult to find an exact equation for the asymptotic
coefficients $\alpha_k$.  As we said, one of them is zero, $(L-1)/2$
are positive, and $(L-1)/2$ are opposite to the positive ones.
Expanding the SBA equations at large $\lambda$ is a bit tricky but
straightforward. We obtain the following equation determining the
positive $\alpha_k>0$
\ba
\label{eq:super}
\exp\left(\frac{i}{2\pi}\,\alpha_k\sum_{\alpha_j>0} \alpha_j\right) &=& \mathop{\prod_{\alpha_j>0}}_{j\neq k} 
\frac{-\frac{i}{2}(\alpha_k-\alpha_j)+2\pi(\alpha_k^{-1}+\alpha_j^{-1})}{\frac{i}{2}(\alpha_k-\alpha_j)+2\pi(\alpha_k^{-1}+\alpha_j^{-1})}\,\times  \nonumber \\
&\times&
\left|\frac{\frac{i}{2}(\alpha_k-\alpha_j)+2\pi(\alpha_k^{-1}+\alpha_j^{-1})}{-\frac{i}{2}(\alpha_k+\alpha_j)+2\pi(\alpha_k^{-1}+\alpha_j^{-1})}\right|^{
4i(h(\alpha_k)-h(\alpha_j))} \\
h(\alpha) &=& \frac{\pi}{\alpha^2}-\frac{\alpha^2}{16\pi}. \nonumber
\ea
where in the l.h.s. the sum includes the case $j=k$. 

Although Eq.~(\ref{eq:super}) is rather complicated, it can be solved
numerically without difficulties, at least starting from the numerical
$p$ obtained at a reasonably large $\lambda$.  As an example, at $L=5,
7, 9$ we find the following (numerical) solutions
\be
\begin{array}{ccc}
L=5 && \\
\hline
\alpha_1  &=& 2.9213116645, \\
\alpha_2  &=& 7.9614845209,
\end{array}
\quad
\begin{array}{ccc}
L=7 && \\
\hline
\alpha_1 &=& 2.1234902933,\\ 
\alpha_2 &=& 4.0857786417, \\  
\alpha_3 &=& 9.1813290270, 
\end{array}
\quad
\begin{array}{ccc}
L=9 && \\
\hline
\alpha_1 &=& 1.6905819725, \\ 
\alpha_2 &=& 3.1495210704,\\ 
\alpha_3 &=& 4.8259074880,\\ 
\alpha_4 &=& 10.203166001 
\end{array}
\ee
These values are in perfect agreement with the numerical solution of
the string Bethe Ansatz equations as illustrated in
Fig.~(\ref{fig:SBAcheck}).

Actually, if we are interested in the asymptotic form of $\Delta^{\rm
string}$, we do not need the full information encoded in the
$\alpha_k$, but just their sum. Indeed,
\be
\frac{\Delta^{\rm string}}{L} \sim c_L\,\lambda^{1/4},
\ee
where
\be
c_L = \frac{1}{2\,L\,\pi}\sum_{k=1}^L \left|\alpha_k \right| = \frac{1}{L\pi}\sum_{\alpha_k>0} \alpha_k.
\ee
We now take the product of Eqs.~(\ref{eq:super}). The right hand sides
cancel perfectly. Evaluating the product of the left hand sides we
obtain
\be
\label{eq:miracle}
\exp\left[\frac{i}{2\pi}\left(\sum_{\alpha_j>0} \alpha_j\right)^2\right] = 1,\quad\Longrightarrow\quad \left(\sum_{\alpha_j>0} \alpha_j\right)^2 = (2\pi)^2\,N_L,
\ee
where $N_L\in\mathbb{N}$. These integers can be determined by solving
Eqs.~(\ref{eq:super}) at a certain $L$. The starting point for
Newton's algorithm is taken from the solution of the SBA equations at
a reasonably large $\lambda$.  Given the solution for the $\alpha_k$
coefficients, we can compute $N_L$. Indeed, the solution of
Eqs.~(\ref{eq:super}) can be accomplished easily with an arbitrarily
high number of digits and the identification of the integer $N_L$ is
totally straightforward and unambiguous.

\medskip
The first values of $N_L$ are 
\be
\begin{array}{cccccccccccc}
L   &:& 5 & 7 & 9   & 11 & 13 & 15 & 17 & 19 & 21 & \dots\\
N_L &:& 3 & 6 &  10 & 15 & 21 & 28 & 36 & 45 & 55 & \dots
\end{array}
\ee
The following simple formula holds
\be
N_L = \frac{1}{8}(L^2-1),
\ee
leading to the prediction
\be
c_L = \frac{1}{L\pi}\sum_{\alpha_k>0} \alpha_k = \frac{1}{L\pi} \,2\pi\sqrt{N_L} = \frac{1}{\sqrt{2}}\left(1-\frac{1}{L^2}\right)^{1/2}.
\ee
Now, we do not observe any particular $\mathbb{Z}_3$ structure. The
values of $c_L$ are perfectly smooth as $L$ increases. As a further
check of this analytical expression, we show in
Fig.~(\ref{fig:stringcfit}) the fit of $c_\infty$ with a simple
quadratic polynomial in $1/L$. There is perfect agreement with the
prediction
\be
c_\infty = \frac{1}{\sqrt{2}}.
\ee
The numerical solution of the SBA equations and the resummation in the
stable region $g^2\lesssim 10$ are in perfect agreement.

\medskip
Of course, the appearance of the integer $N_L$ in the asymptotic form
of $\Delta$ at large $\lambda$ is not surprising. Indeed, this is a
general feature of the SBA equations as discussed
in~\cite{Arutyunov:2004vx}. If the Bethe momenta vanish like
$\lambda^{-1/4}$ at large $\lambda$, then the asymptotic form of
$\Delta$ is
\be
\Delta \sim 2\,\sqrt{n}\,\lambda^{1/4},
\ee
where $n$ is a sum of mode numbers. This is the celebrated string
prediction of~\cite{Gubser:1998bc} where $n$ is the level of a massive
string state. The calculation that we have described identifies the
precise value of $n \equiv N_L$ for the state dual to the highest
operator in the $\mathfrak{su}(1|1)$ sector.

\section{The gauge Bethe Ansatz equations for the AF state in the $\mathfrak{su}(2)$ sector}
\label{sec:GBA2}

It is straightforward to extend the analysis to the highest state in
the $\mathfrak{su}(2)$ sector. The gauge Bethe Ansatz equations read
\be
e^{i\,L\,p_k} = \prod_{j\neq k}
\frac{x^+(p_k)-x^-(p_j)}{x^-(p_k)-x^+(p_j)}\,
\frac{\displaystyle 1-\frac{g^2}{2\,x^+(p_k)\,x^-(p_j)}}
{\displaystyle 1-\frac{g^2}{2\,x^-(p_k)\,x^+(p_j)}},\qquad k = 1, \dots, L,
\ee
Now, we consider $L\in 2\,\mathbb{N}$.

At $g=0$ the above equations reduce to those of the Heisenberg
model. The Bethe momenta $p_k$ are non-trivial and can be determined
numerically at each $L$ as follows. At $g=0$ we have $u_k = \frac 1 2
\cot\frac{p_k}{2}$.  The variables $u_k$ can be determined by solving
({\em e.g.} iteratively)
\be
2\pi\,J_k =  2 \sum_{j=1}^L \arctan(u_k-u_j) -4\,L \arctan(2\,u_k),
\ee
where the Bethe quantum numbers $\{J_k\}$ for the AF state are 
\be
\{J_k\} = \left\{-\frac{L-1}{2}, -\frac{L-3}{2}, \dots, \frac{L-3}{2}, \frac{L-1}{2}\right\}.
\ee
Here, it is not useful to compute the weak coupling expansion of
$\Delta$. Indeed, due to the appearance of the non-trivial one-loop
Bethe roots, all the coefficients of the expansion depend on $L$.

On the other hand, we can integrate numerically the equations. The
result is not surprising and could be expected on the basis of the
Hubbard model solution at finite $L$ discussed
in~\cite{Beccaria:2006aw}.  At large $\lambda$, all Bethe momenta flow
to constant values $p_k\to \overline p_k$ given by
\be
\{\overline p_1, \dots, \overline p_L\} = \left\{\pm\frac{\pi}{L}, \pm\frac{3\pi}{L}, \pm\frac{5\pi}{L}, \dots, \pm\frac{L-1}{L}\pi\right\}.
\ee
Hence, the asymptotic form of the anomalous dimension is 
\ba
\frac{\Delta}{2L} &\sim& \frac{1}{2L}\frac{\sqrt{\lambda}}{\pi}\sum_{\overline p}\left|\sin\frac{\overline p}{2}\right| = \\
&=& 
\frac{\sqrt{\lambda}}{L\,\pi}\sum_{\overline p>0}\sin\frac{\overline p}{2} = 
\frac{\sqrt{\lambda}}{L\,\pi}\sum_{s=0}^{L/2-1} \sin\frac{(2s+1)\pi}{2L} = \frac{\sqrt{\lambda}}{2\,\pi\,L\,\sin\frac{\pi}{2L}}
\ea
In particular, taking $L\to \infty$ we find 
\be
\lim_{L\to\infty}\frac{\Delta}{2 L} = \frac{\sqrt\lambda}{\pi^2} + \dots.
\ee
The factor $2L$ in the scaled anomalous dimension is the correct one,
{\em i.e.} the length of the associated lattice model.  Indeed, in the
$\mathfrak{su}(2)$ sector, the spin zero cyclic state associated with
the AF state with $L$ Bethe momenta has $2L$ spins, $L$ with spin up
and $L$ with spin down.

We remark that the above leading term is obtained both from the GBA
equations and from the Hubbard model.  It is quite interesting to see
what happens at the level of string Bethe Ansatz equations.

\section{The string Bethe Ansatz equations for the AF state in the $\mathfrak{su}(2)$ sector}
\label{sec:SBA2}

The string Bethe Ansatz equations in the $\mathfrak{su}(2)$ sector are
modified by the same universal dressing factor we introduced in the
$\mathfrak{su}(1|1)$ sector. Thus, they read
\be
e^{i\,L\,p_k} = \prod_{j\neq k}
\frac{x^+(p_k)-x^-(p_j)}{x^-(p_k)-x^+(p_j)}\,
\frac{\displaystyle 1-\frac{g^2}{2\,x^+(p_k)\,x^-(p_j)}}
{\displaystyle 1-\frac{g^2}{2\,x^-(p_k)\,x^+(p_j)}}\,e^{i\vartheta(p_k, p_j)},
\ee
where the scattering phase $\vartheta(p_k, p_j)$ has been defined in
Eq.~(\ref{eq:scattering}).

We solve numerically these equations and the outcome is that all Bethe
momenta vanish like $p_k\to \alpha_k\lambda^{-1/4}$ precisely as in
the $\mathfrak{su}(1|1)$ case. Again we can work out an exact equation
for the asymptotic coefficients $\{\alpha_k\}$. Taking the limit of
the SBA equation we find the following modified form of
Eq.~(\ref{eq:super})
\ba
\label{eq:supersu2}
\lefteqn{\exp\left(\frac{i}{2\pi}\,\alpha_k\sum_{\alpha_j>0} \alpha_j\right) = } && \nonumber \\
&& \mathop{\prod_{\alpha_j>0}}_{j\neq k} 
\frac{\frac{i}{2}(\alpha_k+\alpha_j)+2\pi(\alpha_k^{-1}-\alpha_j^{-1})}{-\frac{i}{2}(\alpha_k+\alpha_j)+2\pi(\alpha_k^{-1}-\alpha_j^{-1})}\,
\frac{-\frac{i}{2}(\alpha_k-\alpha_j)+2\pi(\alpha_k^{-1}+\alpha_j^{-1})}{\frac{i}{2}(\alpha_k-\alpha_j)+2\pi(\alpha_k^{-1}+\alpha_j^{-1})}\,\times  \nonumber \\
&\times&
\left|\frac{\frac{i}{2}(\alpha_k-\alpha_j)+2\pi(\alpha_k^{-1}+\alpha_j^{-1})}{-\frac{i}{2}(\alpha_k+\alpha_j)+2\pi(\alpha_k^{-1}+\alpha_j^{-1})}\right|^{
4i(h(\alpha_k)-h(\alpha_j))} \\
h(\alpha) &=& \frac{\pi}{\alpha^2}-\frac{\alpha^2}{16\pi}. \nonumber
\ea
where in the l.h.s. the sum includes the case $j=k$. 

As in the $\mathfrak{su}(1|1)$ sector, we can solve numerically this
equation to cross check the numerical solution of the SBA
equations. For instance, at $L=8$ we have four positive vanishing
momenta and the above equation predicts
\ba
\alpha_1 &=&  2.7192199579,\\ 
\alpha_2 &=&  4.1578685742, \nonumber \\ 
\alpha_3 &=&  5.7295537708, \nonumber \\
\alpha_4 &=& 12.5260989258. \nonumber
\ea
The actual comparison with the solution of the SBA equations is shown
in Fig.~(\ref{fig:SBA2check}). To find the asymptotic expression of
the anomalous dimension we can follow the same strategy as we did in
the $\mathfrak{su}(1|1)$ sector. The extra factors in
Eq.~(\ref{eq:supersu2}) also cancel when all the equations are
multiplied together. Hence, we find again the fundamental relation
\be
\left(\sum_{\alpha_j>0} \alpha_j\right)^2 = (2\pi)^2\,N_L,
\ee
with a different sequence $N_L$. Evaluating the solution to
Eq.~(\ref{eq:supersu2}) we find the table
\be
\begin{array}{cccccccccccc}
L   &:& 4 & 6 & 8   & 10 & 12 & 14 & 16 & 18 & 20 & \dots\\
N_L &:& 4 & 9 & 16  & 25 & 36 & 49 & 64 & 81 & 100& \dots
\end{array}
\ee
Hence, the following simple formula holds
\be
N_L = \frac{1}{4}\,L^2
\ee
leading to the prediction
\be
\frac{\Delta^{\rm string}}{2L}\sim c_L\,\lambda^{1/4},
\ee
with 
\be
c_L = \frac{1}{2\,L\,\pi}\sum_{\alpha_k>0} \alpha_k = \frac{1}{2\,L\,\pi}(2\pi)\frac{L}{2} = \frac{1}{2}.
\ee
This is an exact result holding at {\em any} finite $L$. For instance,
it can be checked that it is valid for the $L=8$ solution up to the
quoted accuracy.

\medskip
Again, the integer $N_L$ appears in the asymptotic expression for
$\Delta$ at large $\lambda$ in the form
\be
\Delta\sim 2\,\sqrt{N_L}\,\lambda^{1/4},
\ee 
identifying $N_L$ with the level of the massive string state dual to
the AF operator.

\section{Discussion}
\label{sec:Discussion}

Quantitative tests of AdS/CFT require a perturbative window allowing
reliable calculations on both sides of the correspondence.  Such a
window does not exist for generic sectors of the spectrum, but is
available for semiclassical string states with large quantum numbers.
In these BMN-like limits, a perturbative check of AdS/CFT at weak
effective 't Hooft coupling can be attempted, but is known to fail at
three loops.  This discrepancy can be seen as a limitation to our
capability of capturing the strong quantum dynamics of string theory
beyond the BMN limit, {\em i.e.} the small $\lambda$ regime for states
with fixed finite classical dimension $L$.

On one hand, the GBA equations are effective in computing all-loop
perturbative gauge theory properties, like anomalous dimensions.
However, the implicit order of limits ($L\to\infty$ after $\lambda\to
0$) spoils the agreement with string calculations beyond two loops.
On the other hand, the SBA equations are valid at large $\lambda$, but
already the leading quantum corrections are known to receive important
and non-trivial corrections at small
$\lambda$~\cite{Beisert:2005cw,Schafer-Nameki:2005tn,Schafer-Nameki:2005is,Schafer-Nameki:2006gk}, not yet under full control despite recent
progresses~\cite{Hernandez:2006tk,Freyhult:2006vr}.

At large $\lambda$, the picture is quite different. In principle, the
GBA equations should not be trusted because the wrapping terms cannot
be neglected. Instead, the SBA equations are expected to match string
calculations, including leading quantum corrections. Hence, we can
make predictions on the string side, but we cannot test the AdS/CFT
correspondence.  Actually, this general statement is very
conservative. Specific cases must be treated with care and a matching
between the solutions of the two sets of equations is possible.
Indeed, the structural difference between GBA and SBA equations lies
only in the dressing scattering phase Eq.~(\ref{eq:scattering}) and
the relevance of this extra term should be considered case by case.

Important examples of such exceptions can be found in various BMN-like
limits. For instance, the exact all loop expression of plane wave
string levels reads in the strict BMN limit 
\be 
\Delta_M - J =
\sum_{k=1}^M \sqrt{1+n_k^2\,\lambda'},\qquad \lambda' =
\frac{\lambda}{J^2},\ J\to\infty, \ee
and scales like $(\lambda')^{1/2}$ at large $\lambda'$. This result is
valid both in the framework of GBA~\cite{Beisert:2004hm} and
SBA~\cite{Arutyunov:2004vx} equations.  In this case, the matching of
the two predictions is due to the fact that the impurities are fixed
in number and their diluteness prevents scattering effects in the
thermodynamical limit.  A related example is the Hofman-Maldacena
limit which also displays asymptotic $(\lambda')^{1/2}$ scaling laws
closely related to the BMN
case~\cite{Hofman:2006xt,Dorey:2006dq,Bobev:2006fg,Spradlin:2006wk,Kruczenski:2006pk}.
Again, in the Hofman-Maldacena limit, the dressing factor in the SBA
equations has been shown to decouple, making the prediction from the
GBA exact, at least in the thermodynamical
limit~\cite{Minahan:2006bd}.

These examples should be regarded as exceptions, precisely because the
irrelevance of the dressing phase is not expected to be generic
feature.  A more involved example where an explicit strong coupling
discrepancy appears is the {\em folded string} (FS)
solution~\cite{Frolov:2003xy} in the $\mathfrak{su}(2)$ sector.  The
energy of this solution is a function $\Delta_{\rm FS}(\lambda,
J)$. In the AdS/CFT correspondence, it must be matched with the
anomalous dimension of an operator with $L=2J$ constituent scalar
fields. The scaling operator is well known, at least in the
$L\to\infty$ limit and is the double contour solution of the GBA
equations described in~\cite{FrolovTseytlin}.  It has been studied in
some details also in the Hubbard model
formulation~\cite{Beccaria:2006aw}.  Setting, as usual $\lambda' =
\lambda/J^2$ we have
\be
\lim_{J\to\infty}\frac{\Delta^{\rm string}_{\rm FS}(\lambda'\,J^2, J)}{2J} = f(\lambda'),
\ee
where the function $f(\lambda')$ is explicitely known. The leading
term is obtained by expanding at large $\lambda'$:
\be
\label{FTstrong}
f(\lambda') \sim \frac{1}{\sqrt{2}}\,(\lambda')^{1/4}.
\ee
One can ask whether it is possible to reproduce Eq.~(\ref{FTstrong})
with the GBA equations.  The anomalous dimension $\Delta^{\rm
gauge}_{\rm FS}$ of the double contour solution is known at strong
coupling and finite $J$ in the Hubbard model GBA
equations~\cite{Beccaria:2006aw}. It reads
\be
\frac{\Delta_{\rm FS}^{\rm gauge}(\lambda, J)}{2J} \sim \frac{1}{\pi\sqrt{2}}\,\cos\frac{\pi}{4J}\,(\lambda')^{1/2}
\ \ \stackrel{J\to\infty}{\longrightarrow}\ \ \frac{1}{\pi\sqrt{2}}\,(\lambda')^{1/2}.
\ee
With the usual remarks about the order of limits, we see that the
GBA/SBA equations predict an asymptotic behaviour $\sim
(\lambda')^\nu$ with $\nu = 1/2$ and $1/4$ respectively.  Here the
number of impurities in the folded string solution gets large as
$J\to\infty$ and their finite density makes the role of the dressing
scattering phase non-trivial.

Following these remarks, it seems interesting to look at other
explicit non-trivial examples where the GBA and SBA equations can be
compared in the strong-coupling region. From a slightly different
perspective, one is considering a special class of states and wonders
about the role of the SBA scattering phase. Following this line of
reasoning, this work analyzes the highest state in the
$\mathfrak{su}(1|1)$ and $\mathfrak{su}(2)$ sectors of ${\cal N}=4$
SYM.  We have been able to solve all ambiguities appearing in the
strong coupling expansion of the Bethe Ansatz equations. Our results
have been cross-checked with a resummation technique which is able to
connect smoothly the weak- and strong-coupling regions.  Our main
results have already been summarized in Eqs.~(\ref{result}) and are
repeated below for the reader's advantage.
\be
\begin{array}{ccl}
\displaystyle\frac{\Delta_{\mathfrak{su}(2)}^{\rm gauge}}{2L}  &=& \displaystyle \frac{1}{2\pi L\sin{\displaystyle\frac{\pi}{2L}}}\,\lambda^{1/2},\\ \\
\displaystyle \frac{\Delta_{\mathfrak{su}(1|1)}^{\rm gauge}}{L}  &=& \displaystyle c_L\,\lambda^{1/2}, 
\end{array}
\qquad
\begin{array}{ccl}
\displaystyle \frac{\Delta_{\mathfrak{su}(2)}^{\rm string}}{2L} &=&  \displaystyle\frac{1}{2}\,\lambda^{1/4}, \\ \\
\displaystyle \frac{\Delta_{\mathfrak{su}(1|1)}^{\rm string}}{L} &=& \displaystyle \frac{1}{\sqrt{2}}\left(1-\frac{1}{L^2}\right)\, \lambda^{1/4},
\end{array}
\ee
$$
c_L \to   0.1405(1),\quad \mbox{as}\ L\to\infty.
$$
\medskip

The main outcome of our analysis is the following. For any fixed $L$,
the highest states in the $\mathfrak{su}(2)$ or $\mathfrak{su}(1|1)$
sectors have a large $\lambda$ anomalous dimension scaling like
$\Delta\sim\lambda^\nu$ where $\nu=1/2$ in the GBA equations and
$\nu=1/4$ in the SBA equations. At large $\lambda$ (and $L$) the SBA
equations can be trusted without subtleties.  Hence, the
$\lambda^{1/2}$ scaling predicted by the GBA equations is not a true
feature of the highest states. Their anomalous dimension immediately
scales like $\lambda^{1/4}$ and uniformly in $L$ as soon as they are
treated by the string Bethe Ansatz equations.

We remark that our result is somewhat novel.  Indeed, in the
$\mathfrak{su}(2)$ sector, it is common lore to believe in the
$\sqrt\lambda$ scaling of the AF operator, after its identification
with the dual of the slow-string limit solution
in~\cite{Roiban:2006jt}. From our analysis, we see that it is
certainly possible to force the SBA equations to exhibit
$\lambda^{1/2}$ scaling. However, this must be done by assuming a
large $\lambda$ behavior of the Bethe momenta that is ruled out
by the explicit solution of the equations, at least for the highest
states.

Notice also that it is possible to quantize the superstring equations
of motion after truncation to the $\mathfrak{su}(1|1)$
sector~\cite{Alday:2005jm,Arutyunov:2005hd}. The spectrum contains
{\em long string} solutions with non-vanishing winding $w = \sum_{k}
p_k$ with $\lambda^{1/2}$ scaling.  On the contrary, {\em short
strings} with vanishing winding exhibit the usual $\lambda^{1/4}$
scaling.  The observed symmetry $p\to -p$ of Bethe momenta favors the
$w=0$ option.

In conclusion, apart from the above mentioned special cases, it seems
definitely dangerous to rely on the GBA equations to estimate the
strong coupling limit of general states, as our analysis of the
highest states has shown.  Instead, the full solution of the SBA
equations, even at the discussed semi-analytical level, appears to be
an effective predictive tool.  For instance, our result~\footnote{
$N_L$ is integer in both sectors since it has been derived with $L$
odd (even) in $\mathfrak{su}(1|1)$ ($\mathfrak{su}(2)$).  }
\be
\label{eq:integerstring}
\begin{array}{lclcccl}
\Delta^{\rm string}_{L, \mathfrak{su}(1|1)} &=& \displaystyle 2\sqrt{N_L}\,\lambda^{1/4}, & \qquad & N_L &=& \displaystyle\frac{1}{8}(L^2-1), \\ \\
\Delta^{\rm string}_{L, \mathfrak{su}(2)}   &=& \displaystyle 2\sqrt{N_L}\,\lambda^{1/4}, & \qquad & N_L &=& \displaystyle\frac{1}{4}L^2,
\end{array}
\ee
gives a simple formula for the level of the string state dual to the
highest state in the two compact sectors.

\appendix

\section{The Weniger resummation algorithm}
\label{app:weniger}

Given the power series 
\be
\lim_{L\to\infty}\frac{\Delta}{L} = \sum_{n\ge 0} c_n\,g^{2n},
\ee
we can evaluate the partial sums
\be
s_n = \sum_{k=0}^n c_k.
\ee
From the partial sums, we form the Weniger approximants
\begin{equation}
\label{dWenTr}
{\delta}_n \; = \; 
\frac {\displaystyle
\sum_{j=0}^{n} \; (- 1)^{j} \; {{n} \choose {j}} \;
\frac {(1 + j)_{n-1}} {(1 + n)_{n-1}} \;
\frac {s_{j}} {c_{j + 1}} }
{\displaystyle
\sum_{j=0}^{n} \; (- 1)^{j} \; {{n} \choose {j}} \;
\frac {(1 + j)_{n-1}} {(1 + n)_{n-1}} \;
\frac {1} {c_{j + 1}} } \, ,
\end{equation}
where $(a)_m = \Gamma(a + m)/\Gamma(a)$ is the Pochhammer symbol.  If
$g^2$ is beyond the convergence radius, the partial sums do not
converge and oscillate wildly. For better stability we have performed
all calculations in exact arithmetics. If the Weniger algorithms
succeeds in resumming the series, then the Weniger approximants
converge.

\newpage

\vskip 2cm
\FIGURE{\epsfig{file=WenigerConvergence.eps,width=14cm} 
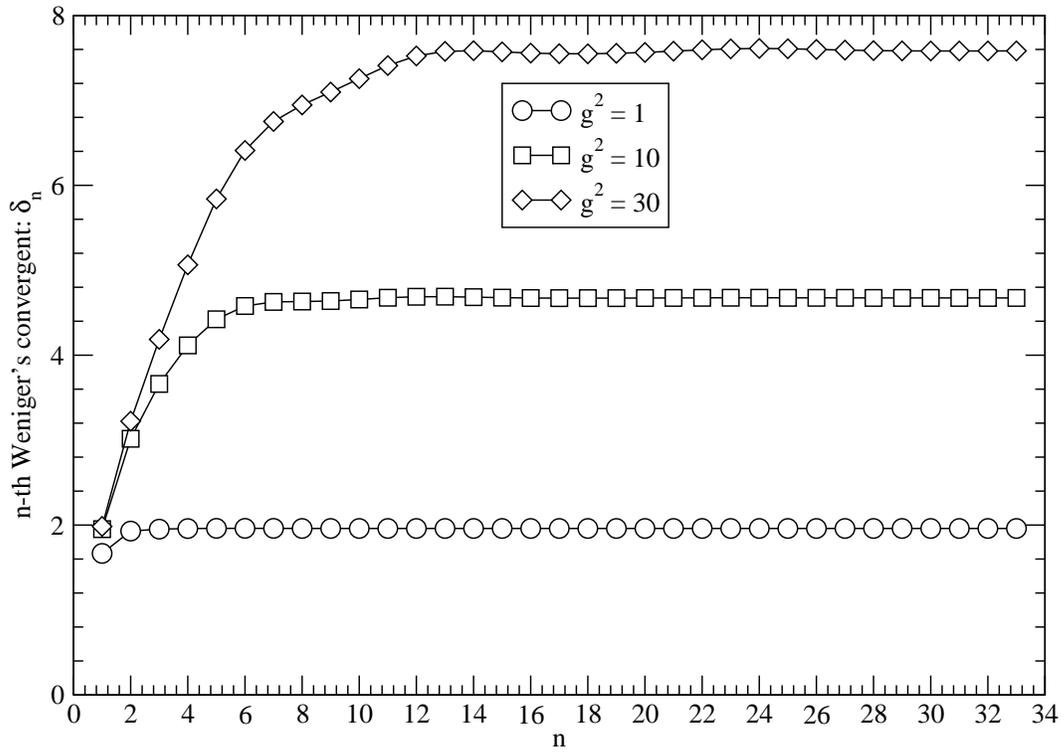
        \caption{Weniger convergents $\delta_n(g^2)$ for the weak coupling expansion of the GBA equations 
         at the three values $g^2 = 1, 10, 30$.}
	\label{fig:wenigerconvergence}}

\vskip 2cm
\FIGURE{\epsfig{file=Weniger.eps,width=14cm} 
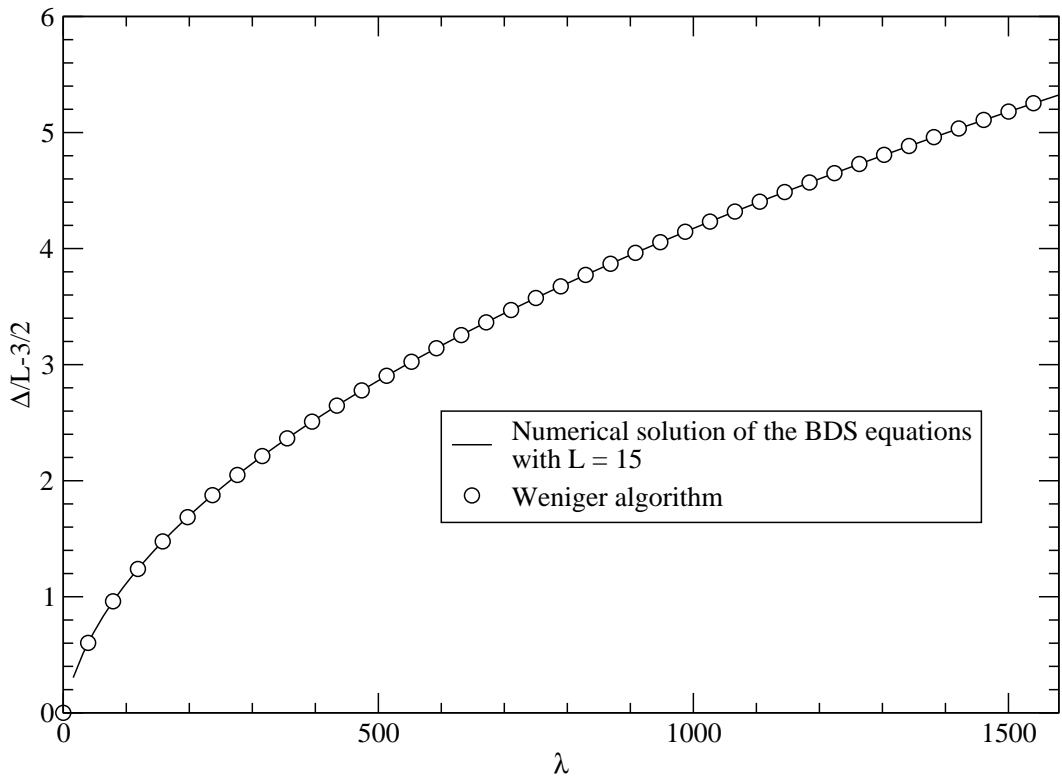
        \caption{Resummation of the weak coupling expansion of the GBA equations. We also show the result from 
        the iterative numerical solution of the same equations at $L=15$.}
	\label{fig:weniger}}

\vskip 2cm
\FIGURE{\epsfig{file=L3.p.eps,width=14cm} 
  \epsfig{file=L5.p.eps,width=14cm} 
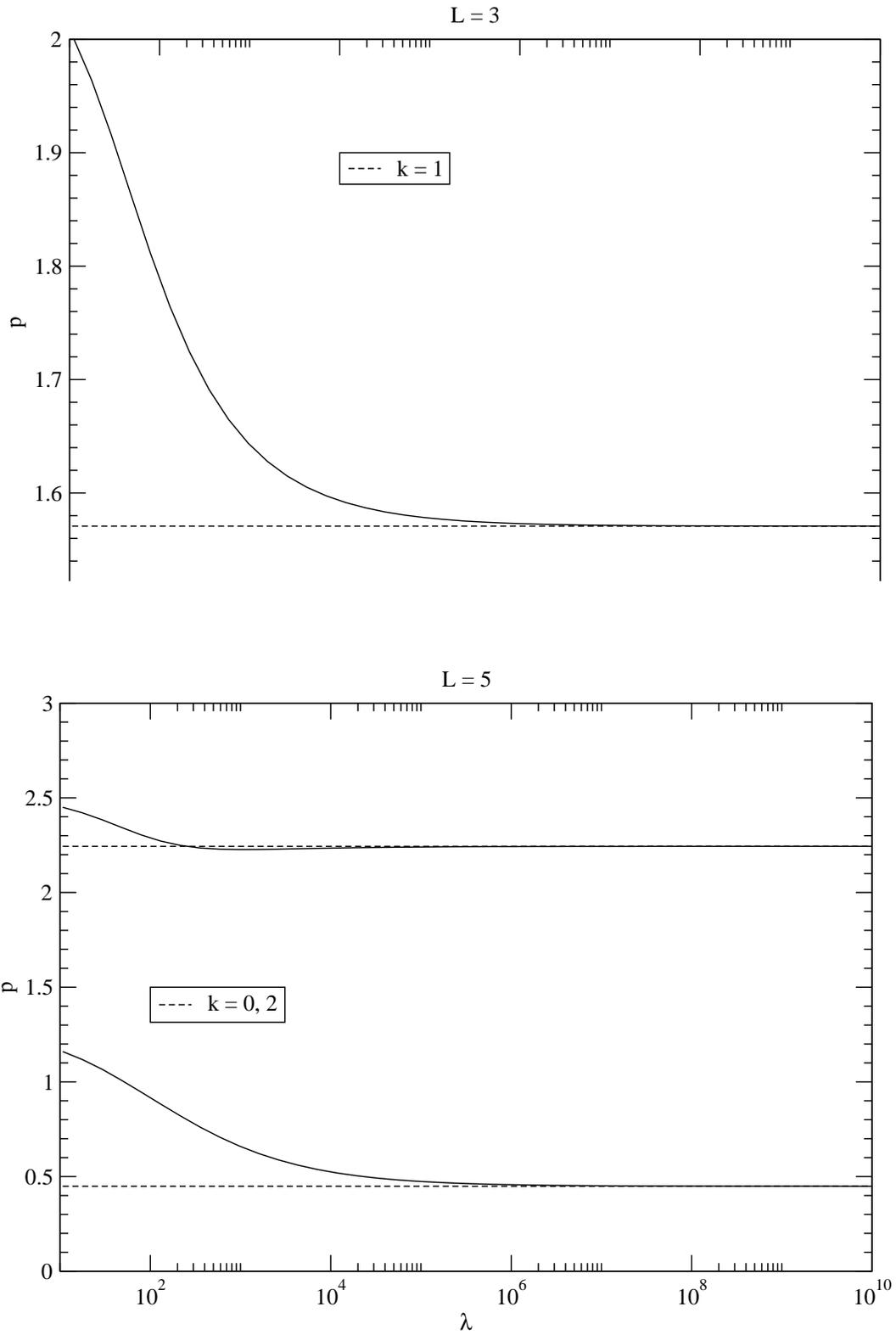
        \caption{Positive Bethe momentum at $L=3,5$. The dashed lines are the analytical prediction.}
	\label{fig:L=3:p}}

\vskip 2cm
\FIGURE{\epsfig{file=L7.p.eps,width=14cm} 
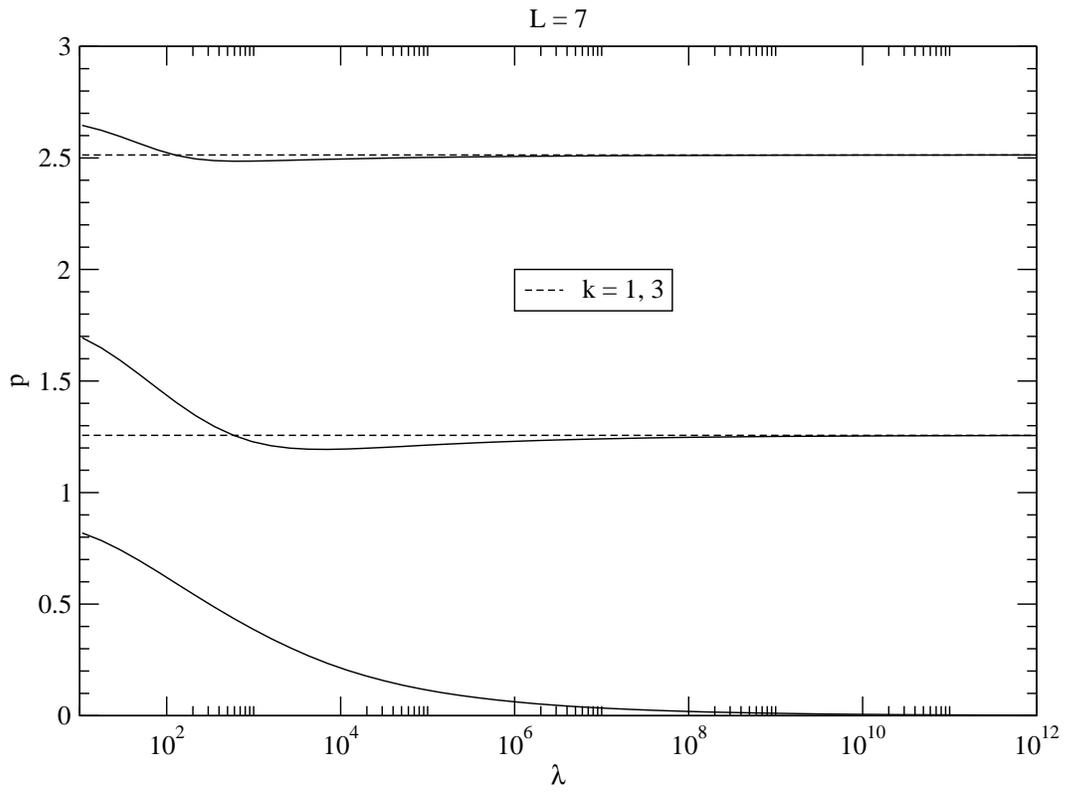
        \caption{Positive and vanishing Bethe momenta at $L=7$. The dashed lines are the analytical predictions.}
	\label{fig:L=7:p}}

\vskip 2cm
\FIGURE{\epsfig{file=L7.pvanishing.eps,width=14cm} 
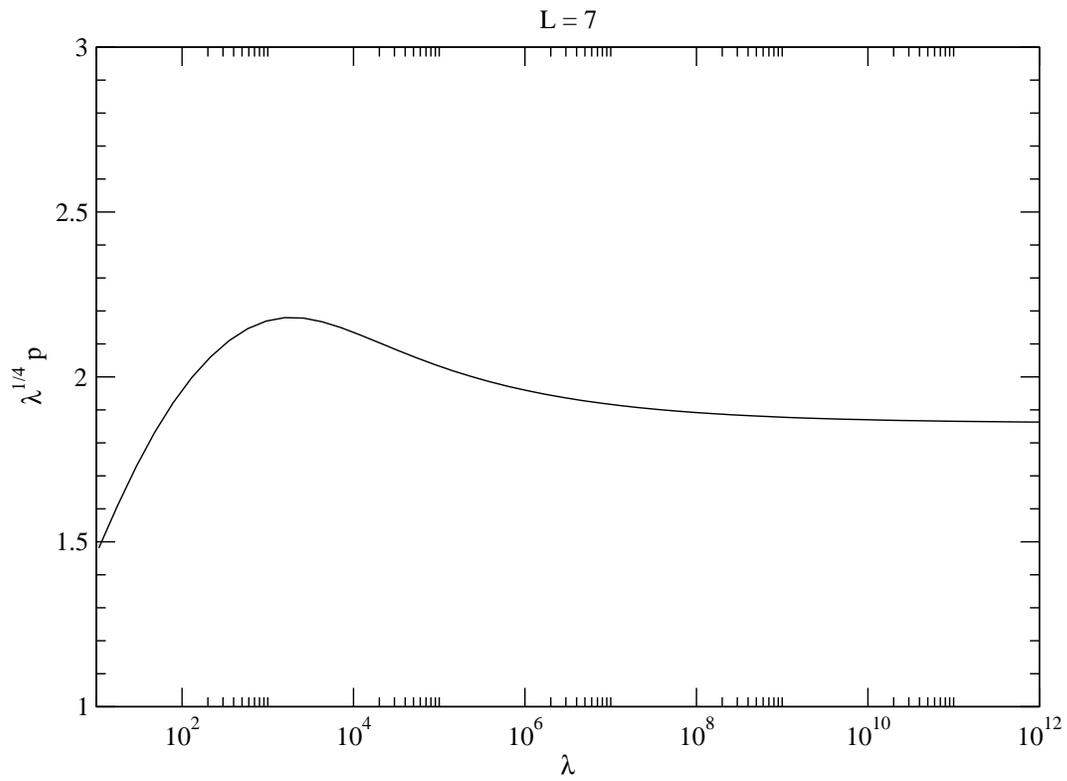
        \caption{$\lambda^{-1/4}$ scaling of the vanishing momentum at $L=7$.}
	\label{fig:L=7:pvanishing}}

\vskip 2cm
\FIGURE{\epsfig{file=L9.p.eps,width=14cm} 
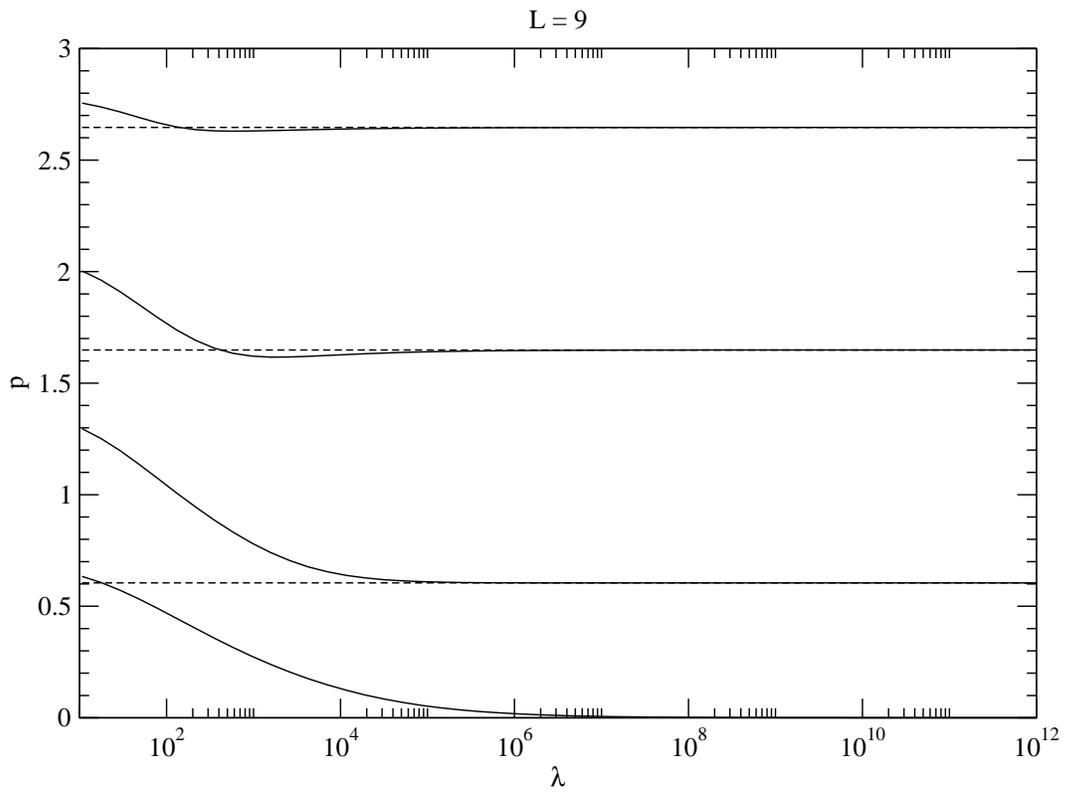
        \caption{Positive and vanishing Bethe momenta at $L=9$. The dashed lines are the analytical predictions.}
	\label{fig:L=9:p}}

\vskip 2cm
\FIGURE{\epsfig{file=L9.pvanishing.eps,width=14cm} 
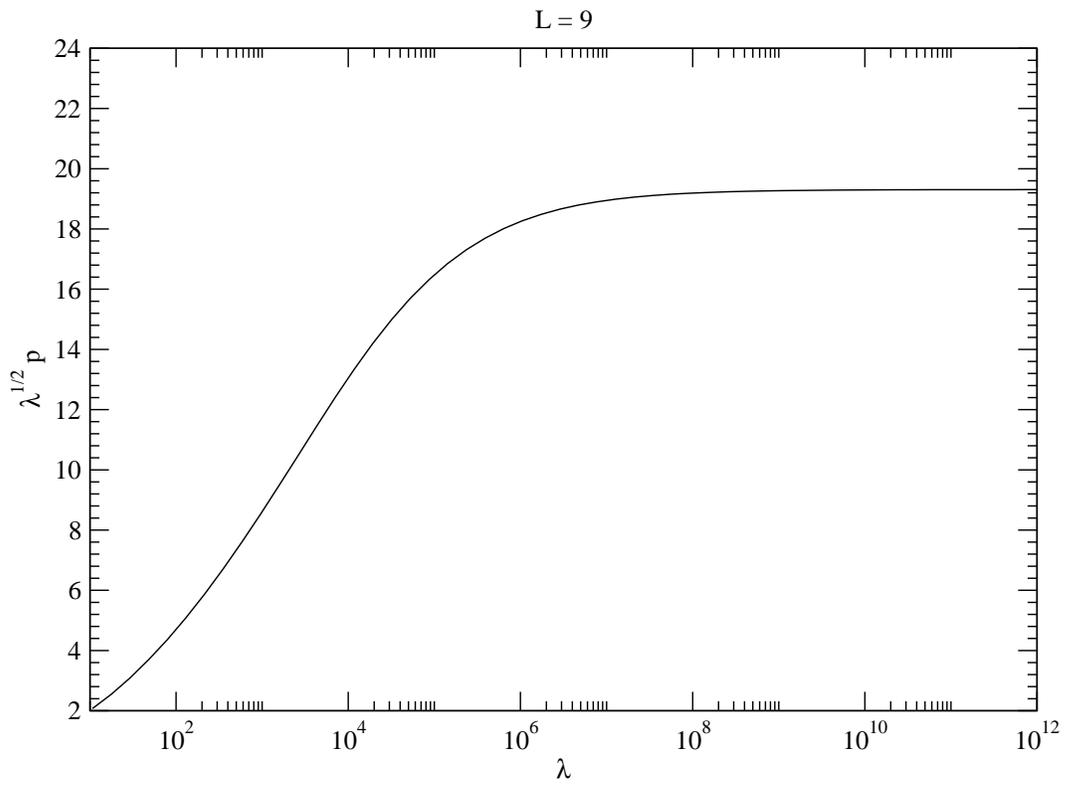
        \caption{$\lambda^{-1/2}$ scaling of the vanishing momentum at $L=9$.}
	\label{fig:L=9:pvanishing}}

\vskip 2cm
\FIGURE{\epsfig{file=L11.p.eps,width=14cm} 
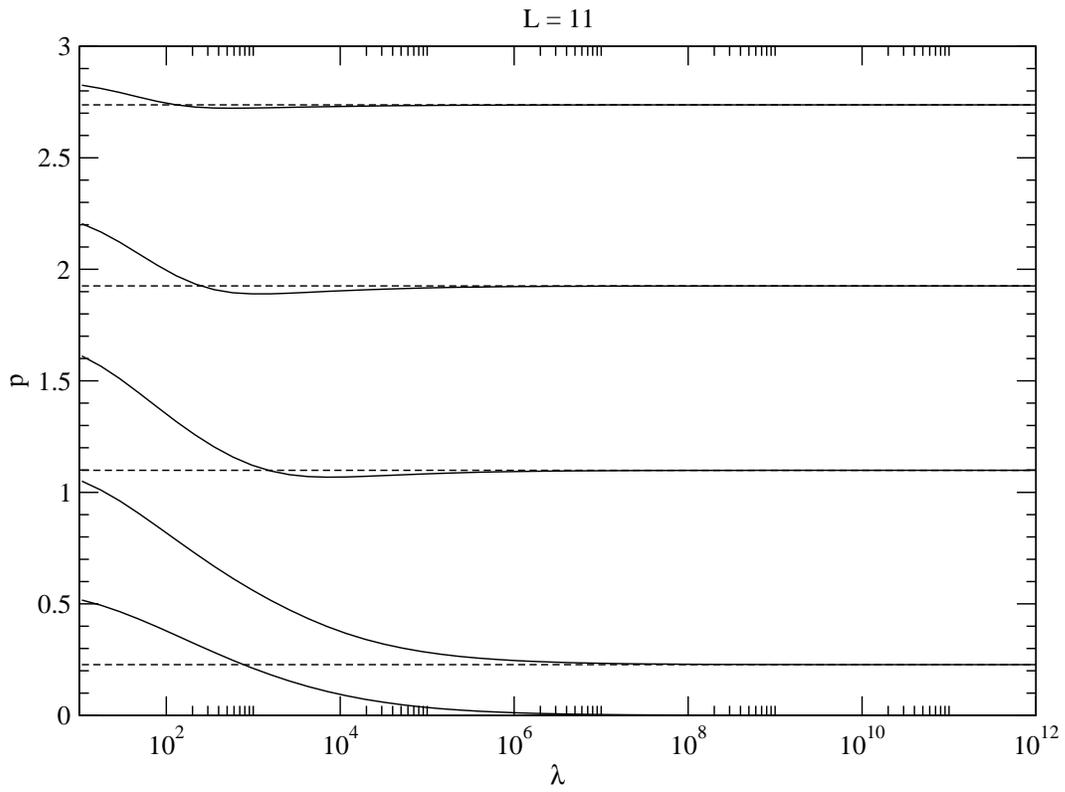
        \caption{Positive and vanishing Bethe momenta at $L=11$. The dashed lines are the analytical predictions.}
	\label{fig:L=11:p}}

\vskip 2cm
\FIGURE{\epsfig{file=L13.p.eps,width=14cm} 
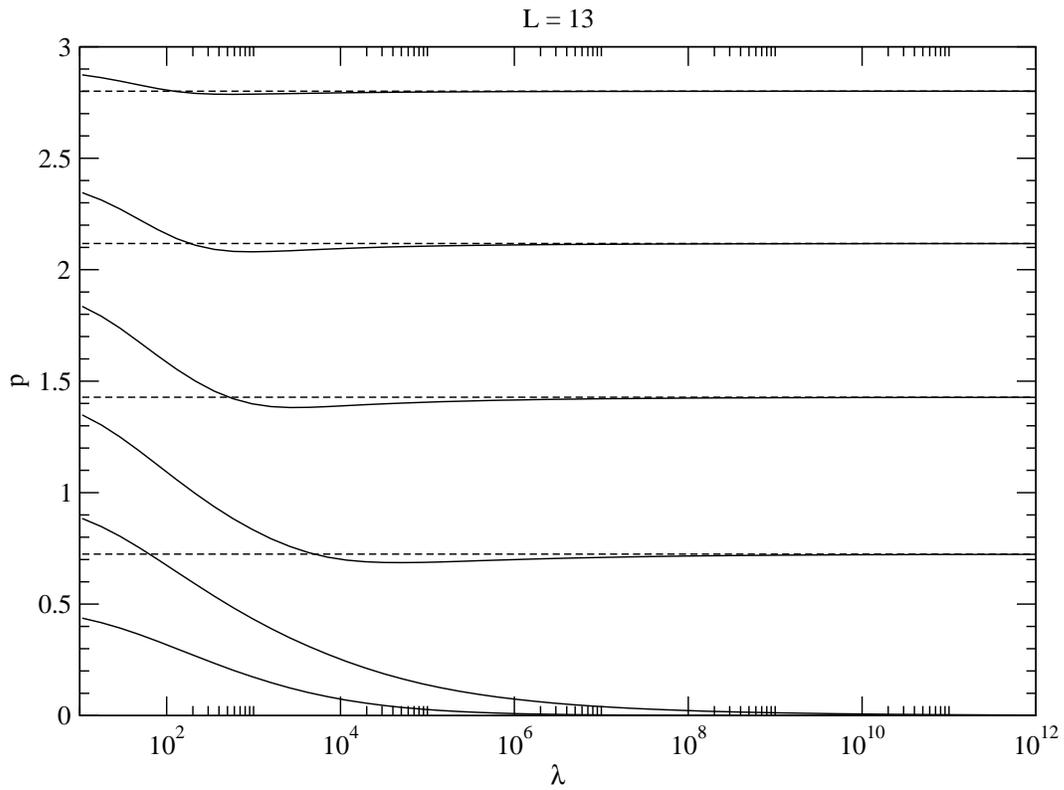
        \caption{Positive and (two) vanishing Bethe momenta at $L=13$. The dashed lines are the analytical predictions.}
	\label{fig:L=13:p}}

\vskip 2cm
\FIGURE{\epsfig{file=L43.p.eps,width=14cm} 
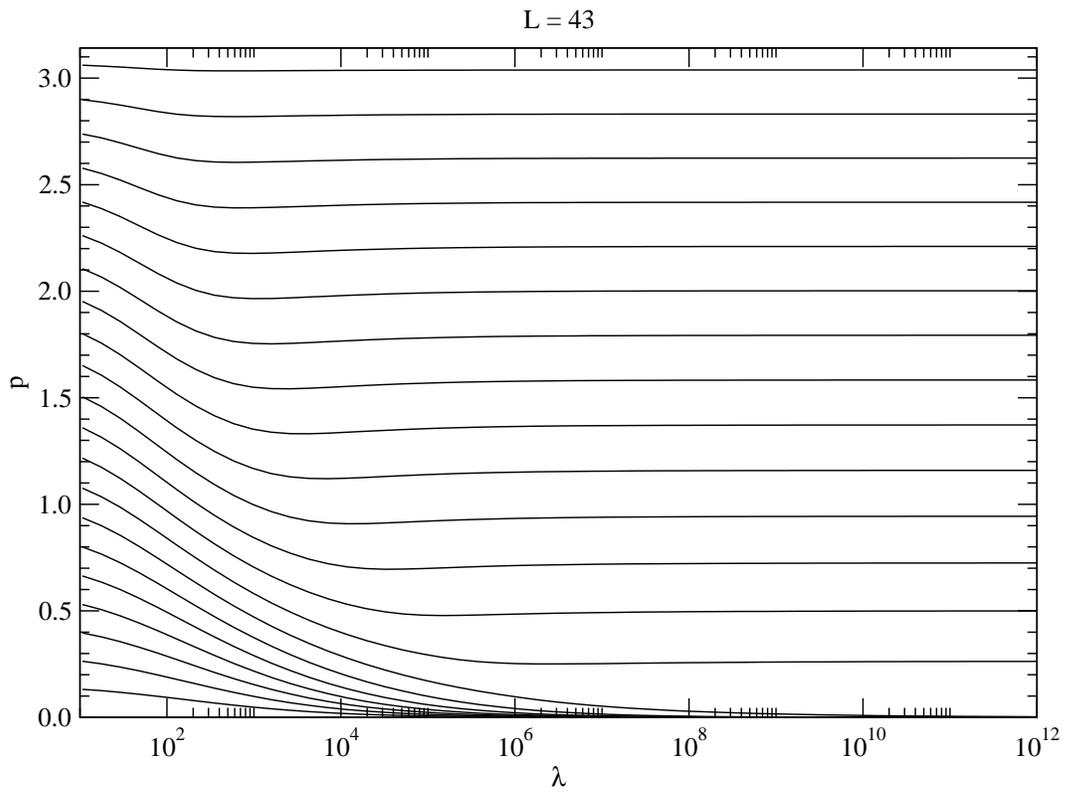
        \caption{Positive and (seven) vanishing Bethe momenta at $L=43$.}
	\label{fig:L=43:p}}

\vskip 2cm
\FIGURE{\epsfig{file=c.eps,width=14cm} 
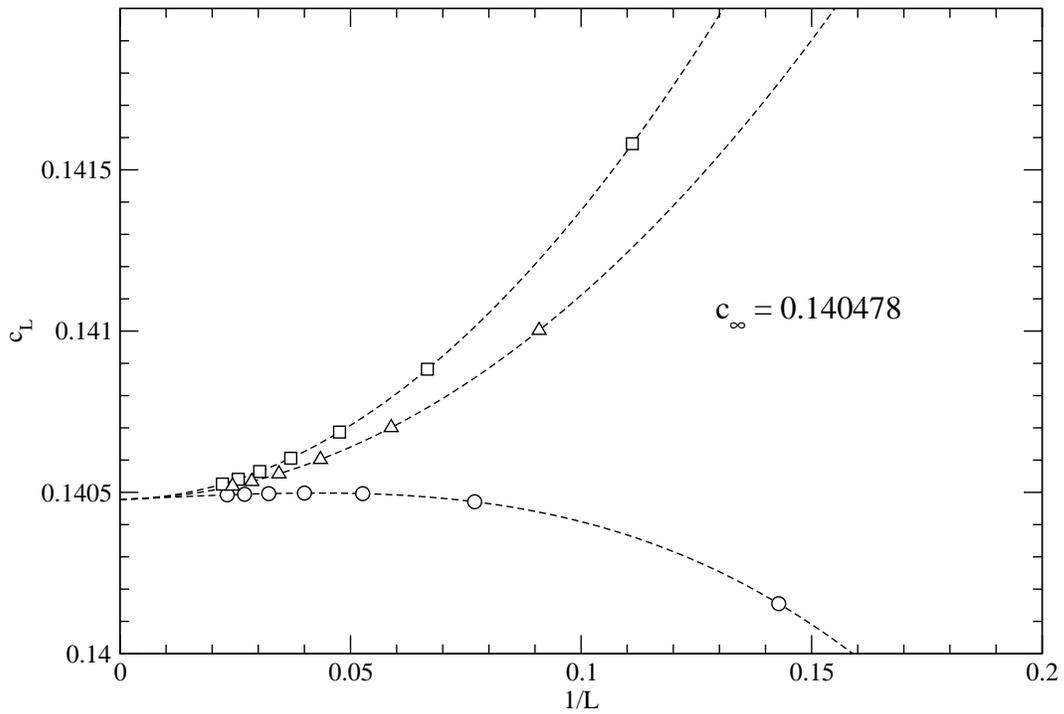
        \caption{Determination of the prefactor $c_\infty$ associated with the leading term in the strong coupling 
expansion of the anomalous dimension predicted by the GBA equations. As explained in the text, the three curves are 
subsequences of lattice length with $((L-1)/2)\, {\rm mod}\, 3 = 0, 1, 2$.}
	\label{fig:cfit}}

\vskip 2cm
\FIGURE{\epsfig{file=comparison.eps,width=14cm} 
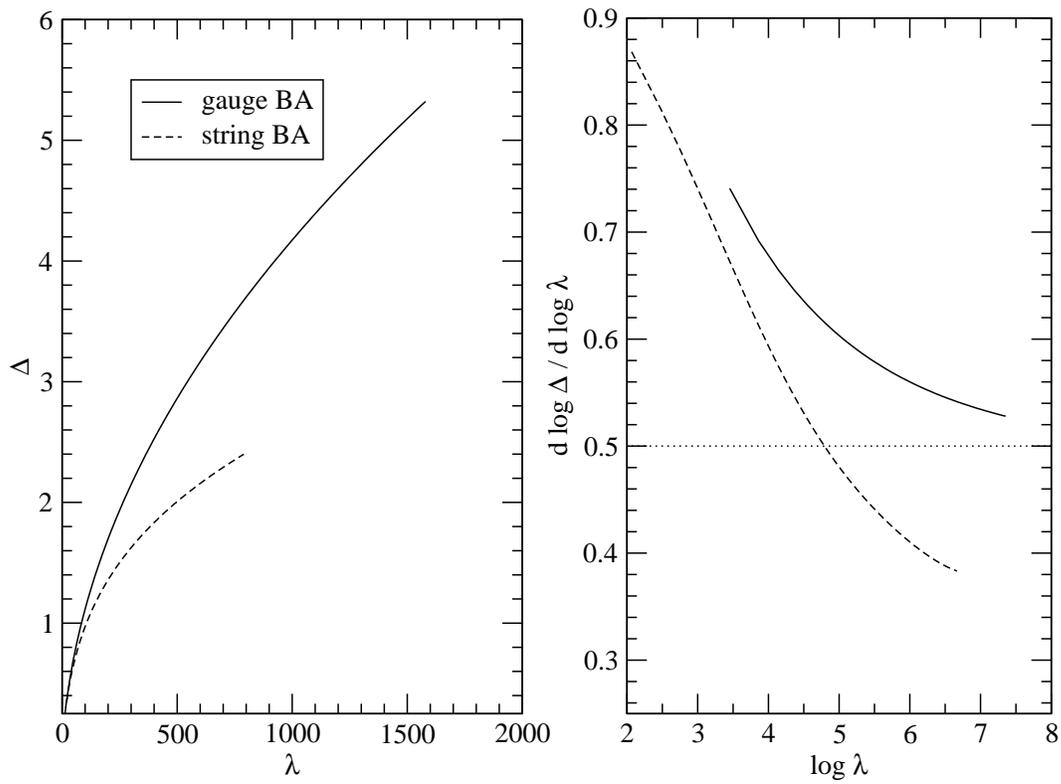
        \caption{Comparison of the gauge/string Bethe Ansatz solutions as predicted by the Weniger resummation algorithm.}
	\label{fig:compare}}

\vskip 2cm
\FIGURE{\epsfig{file=SBAfour.eps,width=14cm} 
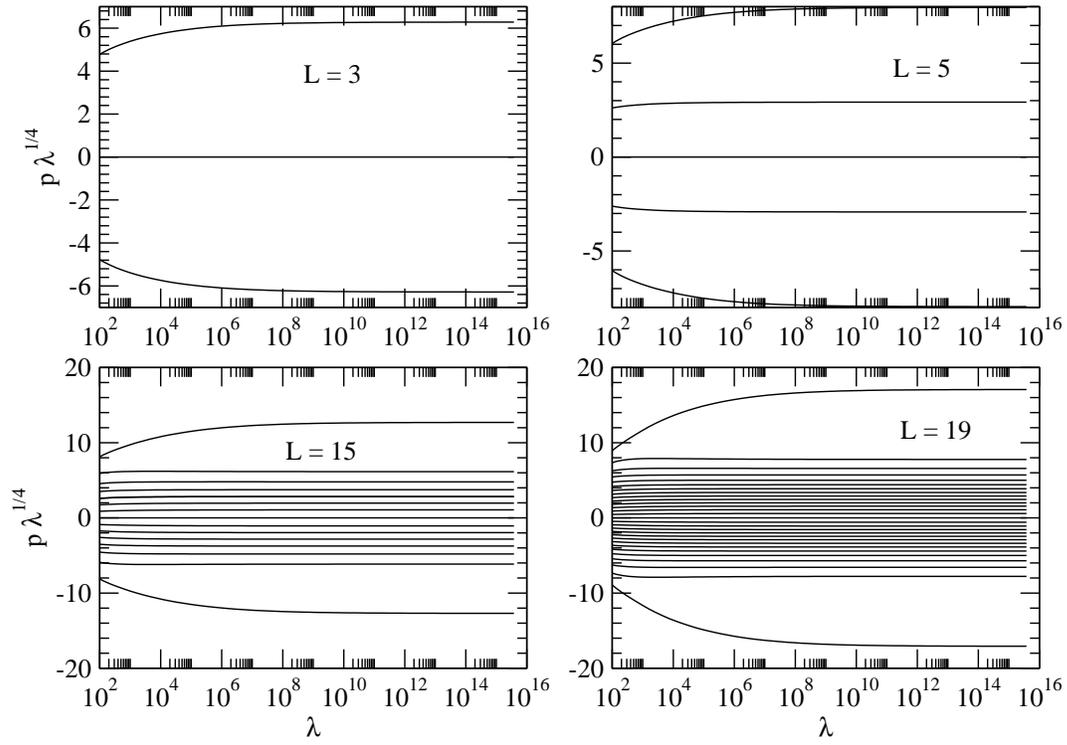
        \caption{Solution of the SBA equations at four $L$. We show a plot of $\lambda^{1/4} p$ for the positive Bethe momenta.}
	\label{fig:SBA}}

\vskip 2cm
\FIGURE{\epsfig{file=SBAcheck.eps,width=14cm} 
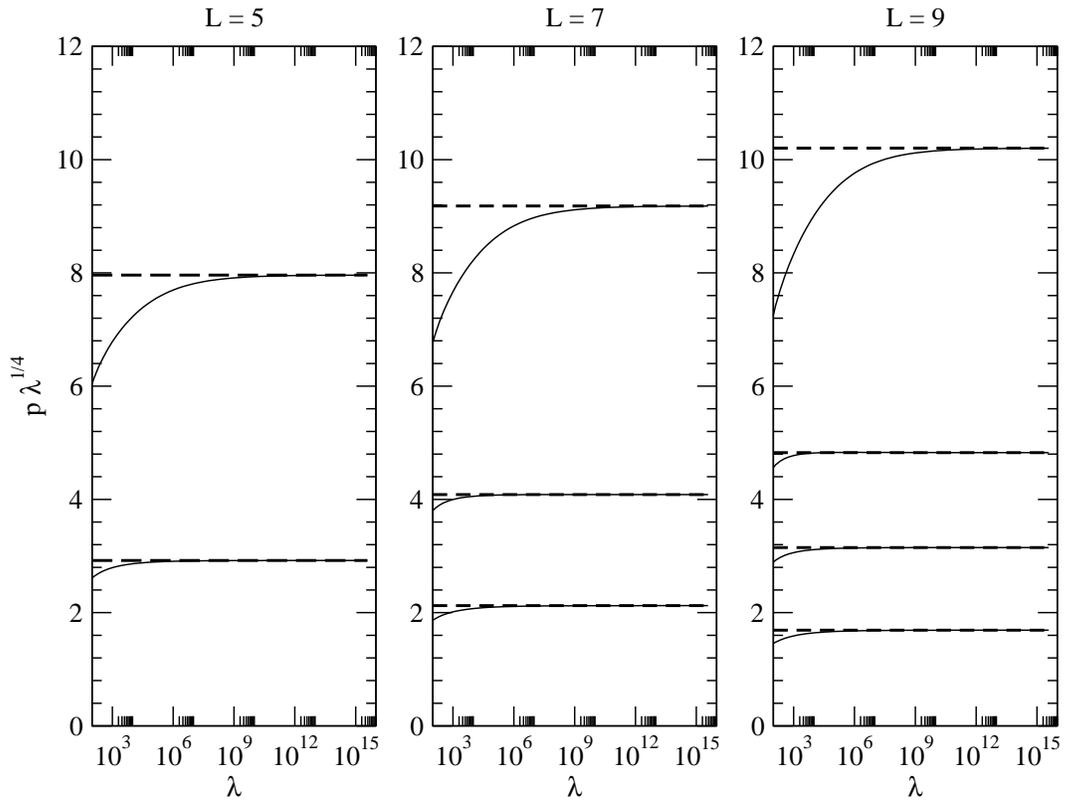
        \caption{Check of the finite $L$ prediction of asymptotic coefficients $\alpha_k$ from the solution of Eq.~(\ref{eq:super}).}
	\label{fig:SBAcheck}}

\vskip 2cm
\FIGURE{\epsfig{file=stringcfit.eps,width=14cm} 
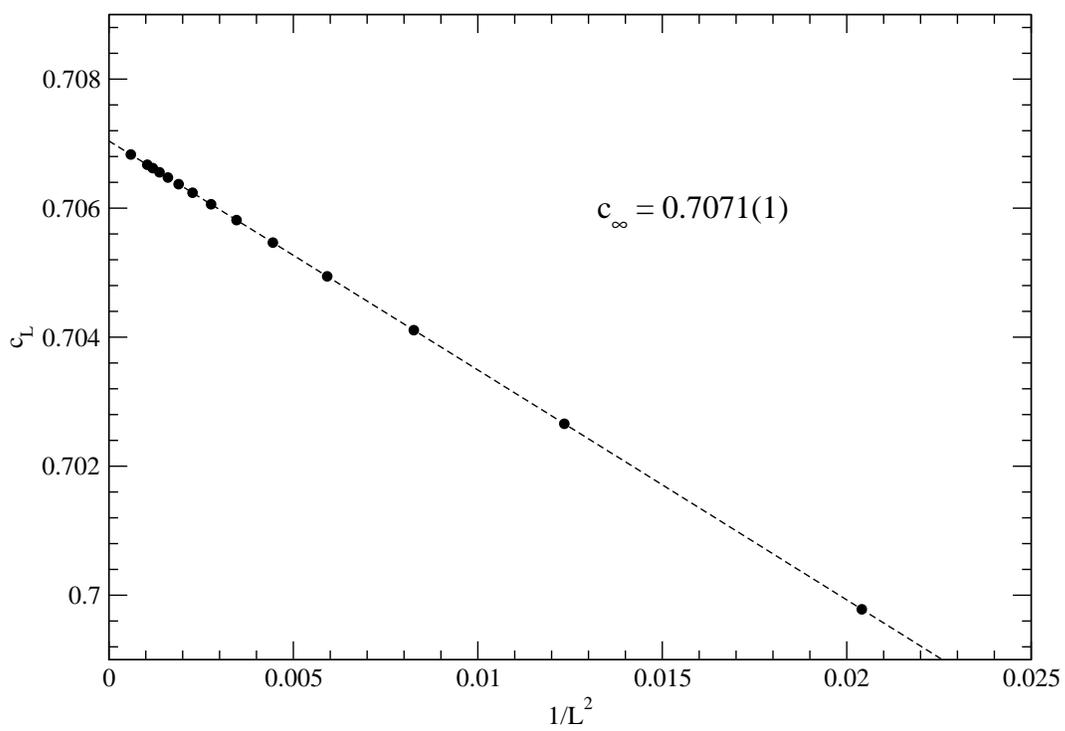
        \caption{Numerical check of the prediction $c_\infty = 1/\sqrt{2}$.}
	\label{fig:stringcfit}}

\vskip 2cm
\FIGURE{\epsfig{file=SBA2check.eps,width=14cm} 
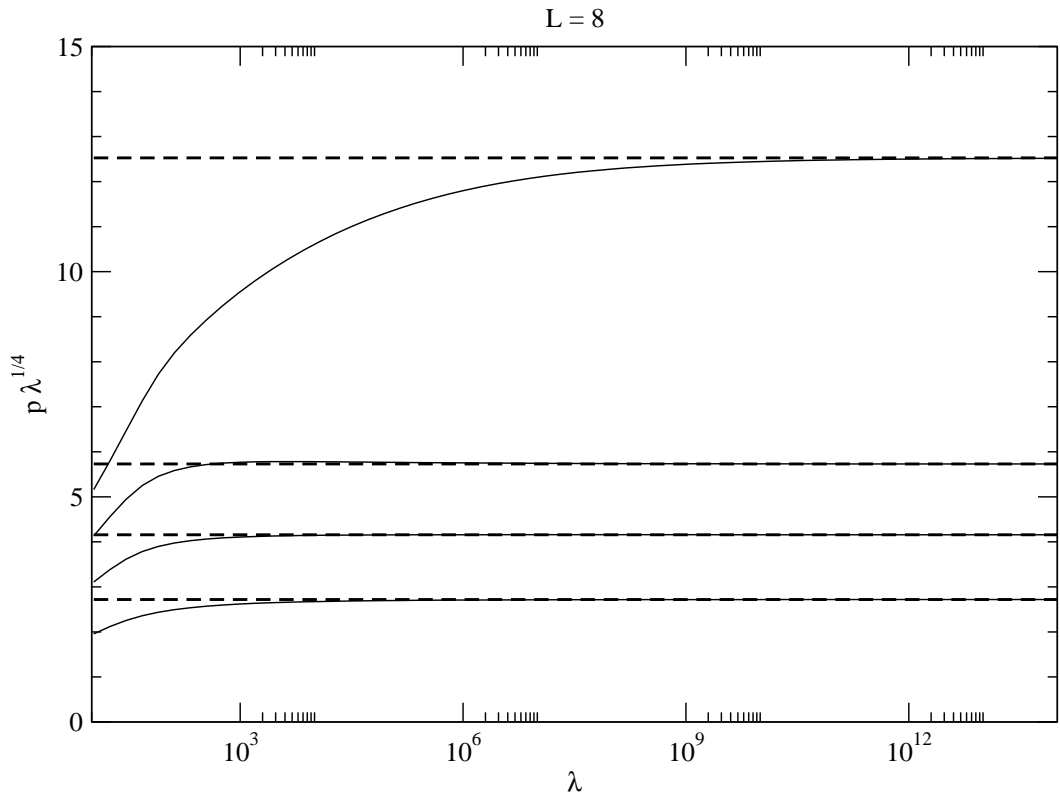
        \caption{Check of the $L=8$ prediction of asymptotic coefficients $\alpha_k$ from the solution of Eq.~(\ref{eq:supersu2}).}
	\label{fig:SBA2check}}

\end{document}